\documentclass[journal]{IEEEtran}
\usepackage{cite}

\ifCLASSINFOpdf
  \usepackage[pdftex]{graphicx}
\else
\fi
\usepackage{amsmath}
\usepackage{amssymb}
\usepackage{amsthm}
\usepackage{bm}
\usepackage{mathalpha}
\usepackage{algorithm}
\usepackage{algpseudocode}
\ifCLASSOPTIONcompsoc
 \usepackage[caption=false,font=normalsize,labelfont=sf,textfont=sf]{subfig}
\else
 \usepackage[caption=false,font=footnotesize]{subfig}
\fi
\usepackage{url}

\usepackage{physics}
\usepackage{booktabs}
\usepackage{threeparttable}
\usepackage{color}
\usepackage{multirow}
\usepackage{stmaryrd}
\SetSymbolFont{stmry}{bold}{U}{stmry}{m}{n}

\theoremstyle{definition}

\newtheorem{remark}{Remark}

\newtheorem{proposition}{Proposition}

\newtheorem*{criterion*}{Criterion}

\newcommand{\figref}[1]{Fig. \ref{#1}}
\newcommand{\tabref}[1]{Table \ref{#1}}
\newcommand{\alref}[1]{Algorithm \ref{#1}}
\newcommand{\appref}[1]{Appendix \ref{#1}}
\newcommand{\secref}[1]{Section \ref{#1}}

\newcommand{\propref}[1]{Proposition \ref{#1}}

\begin{document}
%
\title{A Tensor-Structured Approach to Dynamic Channel Prediction for Massive MIMO Systems with Temporal Non-Stationarity}
%
%
%

\author{Hongwei~Hou,~\IEEEmembership{Graduate~Student~Member,~IEEE},
  Yafei~Wang,~\IEEEmembership{Graduate~Student~Member,~IEEE},
  Yiming~Zhu,~\IEEEmembership{Graduate~Student~Member,~IEEE},
  ~Xinping Yi,~\IEEEmembership{Member,~IEEE},\\
  Wenjin~Wang,~\IEEEmembership{Member,~IEEE},
  Dirk~T.~M.~Slock,~\IEEEmembership{Life~Fellow,~IEEE},
  ~Shi~Jin,~\IEEEmembership{Fellow,~IEEE}
\thanks{}
\thanks{}
\thanks{}}

%
%

\markboth{ }%
{ }
%



\maketitle

\begin{abstract}%
  In moderate- to high-mobility scenarios, channel state information (CSI) varies rapidly and becomes temporally non-stationary, leading to severe performance degradation in the massive multiple-input multiple-output (MIMO) transmissions.
  To address this issue, we propose a tensor-structured approach to dynamic channel prediction (TS-DCP) for massive MIMO systems with temporal non-stationarity, exploiting both dual-timescale and cross-domain correlations.
  Specifically, due to inherent spatial consistency, non-stationary channels over long-timescales can be approximated as stationary on short-timescales, decoupling complicated temporal correlations into more tractable dual-timescale ones.
  To exploit such property, we propose the sliding frame structure composed of multiple pilot orthogonal frequency-division multiplexing (OFDM) symbols, which capture short-timescale correlations within frames by Doppler domain modeling and long-timescale correlations across frames by Markov/autoregressive processes.
  Building on this, we develop the Tucker-based spatial-frequency-temporal domain channel model, incorporating angle-delay-Doppler (ADD) domain channels and factor matrices parameterized by ADD domain grids.
  Furthermore, we model cross-domain correlations of ADD domain channels within each frame, induced by clustered scattering, through the Markov random field and tensor-coupled Gaussian distribution that incorporates high-order neighboring structures.
  Following these probabilistic models, we formulate the TS-DCP problem as variational free energy (VFE) minimization, and unify different inference rules through the structure design of trial beliefs.
  This formulation results in the dual-layer VFE optimization process and yields the online TS-DCP algorithm, where the computational complexity is reduced by exploiting tensor-structured operations.
  Numerical simulations demonstrate the significant superiority of the proposed algorithm over benchmarks in terms of channel prediction performance.
\end{abstract}%

\begin{IEEEkeywords}%
  Massive MIMO, channel prediction, variational inference, tensor representation
\end{IEEEkeywords}%

%
\IEEEpeerreviewmaketitle

\bstctlcite{IEEEexample:BSTcontrol}

\section{Introduction}

\IEEEPARstart{D}{riven} by the demand for higher data rates, massive multiple-input multiple-output (MIMO) and orthogonal frequency-division multiplexing (OFDM) techniques are expected to keep playing vital roles in future communication systems thanks to their significant capacity gain and enhanced spectral efficiency \cite{marzetta2010noncooperative, lu2014overview, bjornson2018massive, sanguinetti2020toward, jin2023massive}. 
In massive MIMO-OFDM systems, precoding and detection rely on the sufficiently accurate channel state information (CSI), making CSI acquisition critical in achieving high performance \cite{liu2018downlink, lee2021downlink, zhu2022ofdm}.

In time division duplex (TDD) systems, the channel reciprocity enables the acquisition of CSI in the uplink for use in the downlink, thus reducing CSI acquisition overhead.
However, large Doppler frequencies and temporal non-stationarity in moderate- to high-mobility scenarios lead to severe channel aging, causing significant performance degradation.
This issue becomes more challenging as future communication systems evolve to the upper mid-band \cite{miao2023sub}, due to the linearity of the Doppler frequency to the carrier frequency.
To address these challenges, channel prediction has emerged as a promising solution, which captures temporal correlations in historical CSI to predict future CSI, thereby combating channel aging \cite{guillaud2004specular}.

\subsection{Previous Works}
In the stationary propagation environments, Doppler domain modeling demonstrates significant superiority among temporal correlation models for channel prediction.
Specifically, \cite{adeogun2015extrapolation} and \cite{adeogun2014parametric} investigate Doppler frequency estimation by spectrum estimation algorithms for narrowband and wideband MIMO systems, respectively.
Leveraging the high angle and delay resolution of massive MIMO-OFDM systems, the angle-delay domain channel estimation is performed, followed by Doppler frequency extraction from effective angle-delay domain channel taps \cite{yin2020addressing, qin2022partial, li2022multi}.
To mitigate the effect of channel estimation error on channel prediction, \cite{zhu2024joint} and \cite{shi2022channel} investigate the joint channel estimation and prediction in the angle-delay-Doppler (ADD) domain.
When it comes to the non-stationary channels, the time-varying Doppler frequencies are approximated by polynomials \cite{wang2023channel},  enabling channel prediction via polynomial Fourier transformation and the orthogonal matching pursuit algorithm.
By treating Doppler frequencies as imperfection parameters, the Doppler variations over time are also captured by autoregressive (AR) processes in \cite{wan2023robust, wan2024two}, which facilitate the exploitation of channel sparsity.

Beyond the Doppler domain modeling, the AR model has also been extensively investigated, especially in non-stationary environments.
Specifically, channel models based on angle and delay domain sparse representation are presented in \cite{ma2019sparse, li2019time, srivastava2020sparse}, where AR processes capture the temporal correlation of sparse domain channels.
As such, sparse Bayesian learning (SBL) and expectation maximization algorithms are employed for channel tracking and AR parameter learning.
By incorporating the AR process of angle domain channels, \cite{srivastava2021sparse} presents the online group SBL algorithm, which exploits the joint sparsity of the angle domain across subcarriers and the structured sparsity within one symbol.
To further explore the inter-subcarrier correlations, angle-delay domain sparse representations are developed in \cite{wu2021channel}, where spatial-temporal AR processes capture the residual temporal correlations of neighboring elements.
Without the perfect channel knowledge, \cite{kim2020massive, wang2024two} adopt spatial domain high-order AR processes to enhance the robustness of channel prediction algorithms in practical systems, while \cite{xu2024learning} employs the deep learning framework.
Due to the continuous evolution of propagation environments, the angle variations over time are also captured by AR processes and tracked by the Kalman filter (KF) \cite{zhao2018time}. 
In \cite{lian2019exploiting, liu2021sparse}, the Markov processes are introduced for the angle and angle-delay domain channel support, capturing the dynamic sparsity.
Inspired by AR processes, \cite{peng2019downlink} employs the Taylor series for temporal correlations of angle-delay domain channels and develops a linear regression-based algorithm.

\subsection{Motivations and Contributions}
Channel prediction in moderate- to high-mobility scenarios is challenging due to intractable temporal correlations arising from dynamic propagation environments.
However, the inherent spatial consistency of channels offers a practical simplification: non-stationary channels over long-timescales can be approximated as stationary on short-timescales \cite{cheng2022channel, huang2022general, hou2024joint}, thus decoupling the complicated temporal correlations into more tractable dual-timescale correlations.
Moreover, the clustered distribution of scatterers induces cross-domain correlations with significant structured patterns in both channel sparsity and power \cite{bian2021general, bian2022novel}.
By capturing these correlations, channel prediction performance in dynamic propagation environments can be significantly enhanced, which motivates our work.

In this paper, we investigate the tensor-structured approach to dynamic channel prediction (TS-DCP) for massive MIMO systems with temporal non-stationarity. The main contributions of this paper are summarized as follows:
\begin{itemize}
  \item To exploit dual-timescale correlations, we propose the sliding frame structure composed of multiple pilot OFDM symbols for channel prediction, characterizing the short- and long-timescale correlations by intra- and inter-frame ones.
  Specifically, we develop the tensor-structured received signal model in the spatial-frequency-temporal (SFT) domain, incorporating ADD domain channels and factor matrices parameterized by ADD domain grids.
  In this representation, the intra-frame correlations are captured by Doppler domain modeling, while the inter-frame correlations of support and power of ADD domain channels are modeled using Markov/AR process.
  \item To capture cross-domain correlations, we introduce the concept of high-order neighbors under the tensor representation, defining the interaction region of cross-domain correlations for each ADD domain grid.
  Building on this, we employ the Markov random fields (MRF) and tensor-coupled Gaussian distributions (TCGD) to model the ADD domain sparse structures and power variations, which are caused by the clustering distribution of scatterers. 
  Specifically, the MRF characterizes the support of ADD domain channels within each frame, while the TCGD models the power variations in the AR process of inter-frame ADD domain channels, providing more accurate representations of practical channels.
  \item Building on the probabilistic models, we formulate the TS-DCP problem as variational free energy (VFE) minimization and present the online approximation to enable real-time prediction.
  By designing the structure of trial beliefs, this problem is interpreted as a dual-layer optimization process, ultimately yielding the online TS-DCP algorithm.
  In this algorithm, different inference rules are unified under the VFE minimization framework, and tensor-structured operations are leveraged to significantly reduce the computational complexity.
  Numerical simulations demonstrate the superior channel prediction performance of the proposed algorithms compared to the state-of-the-art benchmarks.
\end{itemize}

\subsection{Organization and Notations}
\subsubsection{Organization}
The remainder of this paper is organized as follows.
In \secref{sec:system_model}, we develop the SFT domain channel model and its Tucker-based representations with factor matrices parameterized by ADD domain grids.
Based on probabilistic models, the TS-DCP problem is formulated under the VFE minimization framework in \secref{sec:problem_formulation}.
Following this, the online TS-DCP algorithm is developed in \secref{sec:online_TSDCP}, where the inner- and outer-layer are detailed respectively.
Numerical simulations in \secref{sec:simulation_result} demonstrate the superiority of proposed algorithms over benchmarks in terms of the computational complexity and channel prediction performance.
Finally, \secref{sec:conclusion} concludes the paper.

\subsubsection{Standard Notations}
The imaginary unit is represented by $j = \sqrt{-1}$. 
$x$, $\mathbf{x}$, and $\mathbf{X}$ denote scalars, column vectors, and matrices, respectively. 
The transpose, conjugate, and conjugate-transpose operations are represented by the superscripts $(\cdot)^{T}$, $(\cdot)^{\ast}$, and $(\cdot)^{H}$, respectively.
The symbols $\mathbb{C}$ denote the complex number fields.
$[\cdot]_{i_{1}, {\dots}, i_{D}}$ is the $(i_{1}, {\dots}, i_{D})$-th element of $D$-order tensor.
$\mathsf{D}\left\{\cdot\right\}$ and $\mathsf{H}\left\{\cdot\right\}$ denote the Kullback-Leibler (KL) divergence and differential entropy, respectively.
The outer and Hadamard products are represented by the operator $\circ$ and $\odot$, respectively.
$\mathsf{E}\left\{\cdot\right\}$ and $\mathsf{V}\left\{\cdot\right\}$ denote the expectation and variance operators, respectively.
$\mathsf{diag}\{\cdot\}$ and $\mathsf{Re}\{\cdot\}$ denote the diagonal and real part operators, respectively.

\subsubsection{Tensor Notations}
The tensor operations and definitions in this paper align with the counterparts in \cite{kolda2009tensor}. For a $D$-order tensor $\boldsymbol{\mathcal{X}}{\;\in\;}\mathbb{C}^{N_{1}{\times}N_{2}{\times}{\dots}{\times}N_{D}}$, 
the mode-$d$ matrixization $\mathbf{X}_{d}{\;\in\;}\mathbb{C}^{N_{d}{\times}N_{1}{\dots}N_{d-1}N_{d+1}{\dots}N_{D}}$ arranges the mode-$d$ fibers of this tensor, obtained by fixing the index along the $d$-th dimension and varying the others, into its column vectors.
The high-order cyclic shift operations is defined by $[\boldsymbol{\mathcal{X}}^{[\mathbf{r}]}]_{n_{1}, \dots, n_{D}} = [\boldsymbol{\mathcal{X}}]_{\bar{n}_{1}, \dots, \bar{n}_{D}}$ with $\bar{n}_{d} = (n_{d}+r_{d}) \mod {N_{d}}$. 
Given the tensors $\boldsymbol{\mathcal{X}}, \boldsymbol{\mathcal{Y}}$ with the same size, the inner product is defined as
\begin{equation}
  \langle \boldsymbol{\mathcal{X}}, \boldsymbol{\mathcal{Y}} \rangle = \sum_{n_{1}}\sum_{n_{2}}{\dots}\sum_{n_{D}} [\boldsymbol{\mathcal{X}}]_{n_{1}, n_{2}, {\dots}, n_{D}}[\boldsymbol{\mathcal{Y}}]_{n_{1}, n_{2}, {\dots}, n_{D}}^{\ast},
\end{equation}
which is also the high-order extension of matrix inner product, and we define $\|\boldsymbol{\mathcal{X}}\|_{F} = \sqrt{\langle \boldsymbol{\mathcal{X}}, \boldsymbol{\mathcal{X}} \rangle}$ as the high-order extension of Frobenius norm. 
The mode-$d$ tensor-matrix multiplication of tensor $\boldsymbol{\mathcal{X}}$ and matrix $\mathbf{U}_{d}{\;\in\;}\mathbb{C}^{K_{d}{\times}N_{d}}$ is denoted as $\boldsymbol{\mathcal{Y}} = \boldsymbol{\mathcal{X}}{\;\times_{d}\;}\mathbf{U}_{d}$, expressed as
\begin{equation}
  [\boldsymbol{\mathcal{Y}}]_{n_{1},{\dots}, n_{d-1}, k_{d}, n_{d+1},{\dots},n_{D}} = \sum_{n_{d}}[\mathbf{U}]_{k_{d},n_{d}}[\boldsymbol{\mathcal{X}}]_{n_{1},{\dots},n_{D}},
\end{equation}
and we have $\mathbf{Y}_{d} = \mathbf{U}_{d}\mathbf{X}_{d}$. $\mathsf{CN}( \boldsymbol{\mathcal{X}}; \boldsymbol{\mathcal{U}}, \boldsymbol{\mathcal{E}})$ denotes the joint probability density function (PDF) of $\boldsymbol{\mathcal{X}}$, whose entries are independently drawn from circularly symmetric complex Gaussian distributions with element-wise mean $\boldsymbol{\mathcal{U}}$ and variance $\boldsymbol{\mathcal{E}}$.
$\boldsymbol{\mathcal{C}}(x)$ denotes the constant tensor with all elements being $x$. The element-wise function applications to tensors follow the convention that, for any tensor $\boldsymbol{\mathcal{X}}$ and function $f(\cdot)$, $[f(\boldsymbol{\mathcal{X}})]_{n_{1}, \dots, n_{D}} = f([\boldsymbol{\mathcal{X}}]_{n_{1}, \dots, n_{D}})$, where $f{\;\in\;}\{\mathsf{exp}, \mathsf{ln}\}$ denotes the exponential or logarithmic function.

\section{System Model}\label{sec:system_model}
In this paper, we consider the massive MIMO-OFDM system with a base station (BS) equipped with a uniform planar array (UPA) of $N_\text{an} = N_\text{h}{\;\times\;}N_\text{v}$ antennas and single-antenna mobile terminals (MTs), where $N_\text{h}$ and $N_\text{v}$ denote the number of horizontal and vertical antennas, respectively.

\begin{figure}[!t]
  \centering
  \includegraphics[width = \linewidth]{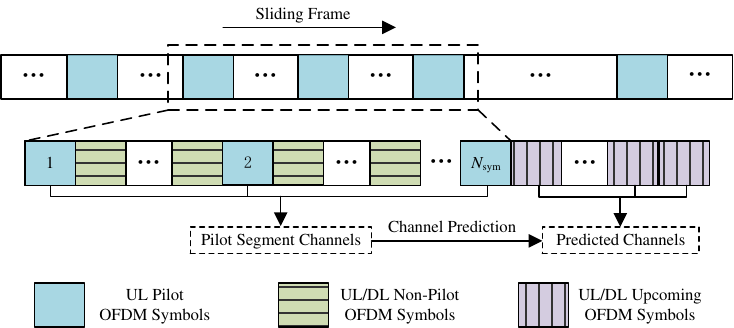}
  \caption{Sliding frame structure for dynamic channel prediction.}
  \label{fig:FrameStructure}
\end{figure}

The system operates in TDD mode under the assumption of perfect calibration, and adopts sounding reference signal as pilots for uplink channel sounding \cite{3gpp38211}. The MTs employ comb-type pilot patterns, with uniform pilot spacing of ${\Delta}\bar{T}$ in the temporal domain and ${\Delta}\bar{f}$ in the frequency domain, respectively.
Since channels exhibit inherent spatial consistency in practical systems, the physical parameters of each path remain unchanged on short-timescale and vary smoothly on long-timescale \cite{cheng2022channel, huang2022general, hou2024joint}. 
To capture this dual-timescale correlation, we introduce the sliding frame structure shown in \figref{fig:FrameStructure}, where each frame contains multiple pilot OFDM symbols. 
Within each frame, intra-frame correlations characterize short-timescale channel variations, while inter-frame correlations capture long-timescale dynamic correlations across frames.
In each frame, the BS collects the received symbols in the pilot segment and predicts channels of all symbols from the last observed pilot symbol to the first upcoming pilot symbol, thus combating channel aging.

\subsection{Channel and Signal Models}
The channel impulse response (CIR) at the $(n_\text{h}, n_\text{v})$-th antenna and the $n_\text{F}$-th frame is expressed as\footnote{Since there is no pilot contamination between MTs, we focus on the typical MT in the cell and ignore the MT index for simplicity of expressions.}
\begin{equation}
  h_{n_\text{F}, n_\text{h}, n_\text{v}}(t, \tau) = \sum_{l=1}^{L_{n_\text{F}}} {g}_{n_\text{F}, l} e^{j2{\pi}t{\nu}_{n_\text{F}, l}}\delta( {\tau} - \tau_{n_\text{F}, n_\text{h}, n_\text{v}, l}),
\end{equation}
where $L_{n_\text{F}}$ denotes the number of paths, $\tau_{n_\text{F}, n_\text{h}, n_\text{v}, l} {\;\triangleq\;} {\tau}_{n_\text{F}, l} + [(n_\text{h} - 1){d}_{\text{h}, n_\text{F}, l} + (n_\text{v} - 1){d}_{\text{v}, n_\text{F}, l}]/c$ denote the propagation delay of the $l$-th path from the MT to the $(n_\text{h}, n_\text{v})$-th under the far-field assumption, ${d}_{\text{h}, n_\text{F}, l}{\;\triangleq\;}d\mathsf{sin}({\psi}_{\text{el}, n_\text{F}, l})\mathsf{sin}({\psi}_{\text{az}, n_\text{F}, l})$ and ${d}_{\text{v}, n_\text{F}, l}{\;\triangleq\;}d\mathsf{cos}({\psi}_{\text{el}, n_\text{F}, l})$ denote the propagation distance difference in horizontal and vertical dimensions of the $l$-th path, respectively, ${g}_{n_\text{F}, l}$, ${\tau}_{n_\text{F}, l}$, ${\nu}_{n_\text{F}, l}$, ${\psi}_{\text{el}, n_\text{F}, l}$, and ${\psi}_{\text{az}, n_\text{F}, l}$ denote the complex gain, delay, Doppler frequency, elevation and azimuth angle of the $l$-th path, respectively, $c$ and $d$ denote the speed of light and inter-antenna spacing, respectively.

By taking the Fourier transform of CIRs and stacking the result in all antennas and pilot resource elements in the $n_\text{F}$-th frame, the SFT domain channel at pilot segments is given as
\begin{align}\label{eq:channel_model_cp}
  \boldsymbol{\mathcal{H}}_{n_\text{F}} = \sum_{l=1}^{L_{n_\text{F}}} g_{n_\text{F}, l} \mathbf{a}_{\text{h}}({\theta}_{n_\text{F}, l}) {\;\circ\;} \mathbf{a}_{\text{v}}({\phi}_{n_\text{F}, l}) {\;\circ\;} \mathbf{b}({\tau}_{n_\text{F}, l}) {\;\circ\;} \mathbf{c}({\nu}_{n_\text{F}, l}),
\end{align}
where $\mathbf{a}_{\text{h}}({\theta})$, $\mathbf{a}_{\text{v}}({\phi})$, $\mathbf{b}(\tau)$ and $\mathbf{c}({\nu})$ denote the horizontal angle, vertical angle, delay and Doppler domain steering vectors, defined by $[\mathbf{a}_{\text{h}}({\theta})]_{n_\text{h}} {\;\triangleq\;} \mathsf{exp}(-j2{\pi}(n_\text{h} - 1){\theta})$, $[\mathbf{a}_{\text{v}}({\phi})]_{n_\text{v}} {\;\triangleq\;} \mathsf{exp}(-j2{\pi}(n_\text{v} - 1){\phi})$, $[\mathbf{b}(\tau)]_{n_\text{sc}}{\;\triangleq\;}\mathsf{exp}(-j2{\pi}(n_\text{sc} - 1){\Delta}\bar{f}{\tau})$ and $[\mathbf{c}({\nu})]_{n_\text{sym}}{\;\triangleq\;}\mathsf{exp}(j2{\pi}(n_\text{sym} - 1){\Delta}\bar{T}{\nu})$, respectively, ${\theta}_{n_\text{F}, l} {\;\triangleq\;} {d}_{\text{h}, n_\text{F}, l} / {\lambda} $ and ${\phi}_{n_\text{F}, l} {\;\triangleq\;} {d}_{\text{v}, n_\text{F}, l} / {\lambda} $ denote the horizontal and vertical directional cosines of the $l$-th path, respectively, and $\lambda$ denotes the wavelength.

After the cyclic prefix removal and OFDM demodulation, the received signal tensor at pilot segments is expressed as
\begin{equation}\label{eq:received_sig}
  \boldsymbol{\mathcal{Y}}_{n_\text{F}} =  \boldsymbol{\mathcal{X}} {\;\odot\;} \boldsymbol{\mathcal{H}}_{n_\text{F}} + \boldsymbol{\mathcal{Z}}_{n_\text{F}},
\end{equation}
where $\boldsymbol{\mathcal{X}}{\;\in\;}\mathbb{C}^{N_\text{h}{\times}N_\text{v}{\times}N_\text{sym}{\times}N_\text{sc}}$, $\boldsymbol{\mathcal{H}}_{n_\text{F}}{\;\in\;}\mathbb{C}^{N_\text{h}{\times}N_\text{v}{\times}N_\text{sym}{\times}N_\text{sc}}$ and $\boldsymbol{\mathcal{Z}}_{n_\text{F}}{\;\in\;}\mathbb{C}^{N_\text{h}{\times}N_\text{v}{\times}N_\text{sym}{\times}N_\text{sc}}$ denote the pilot tensor, the SFT domain channel and the additive white Gaussian noise at pilot segments, respectively. Since the pilot symbols in the uplink training phase are known at BS, we assume that $\boldsymbol{\mathcal{X}}$ is all-one tensor without loss of generality.

\subsection{Tucker-Based Representations}
Due to the limited scattering propagation environments, the SFT domain channels exhibit low-rank structures, which can be exploited by the unstructured tensor decomposition (TD) framework.
However, the uniqueness condition of such framework is typically not satisfied in practical propagation environments. This limitation can be mitigated by leveraging the inherent structure of factor matrices \cite{sorensen2013blind, goulart2015tensor, cohen2017dictionary}.

Based on \eqref{eq:channel_model_cp}, the factor matrices of SFT domain channels are parameterized by ADD domain grids, defined as
\begin{subequations}
  \begin{equation}
    \mathbf{A}_{{\text{h}}}(\tilde{\bm{\theta}}_{n_\text{F}}) = [ \mathbf{a}_\text{h}(\tilde{\theta}_{1, n_\text{F}}), {\dots}, \mathbf{a}_\text{h}(\tilde{\theta}_{K_\text{h}, n_\text{F}}) ] {\;\in\;}\mathbb{C}^{N_\text{h}{\times}K_\text{h}},
  \end{equation}
  \begin{equation}
    \mathbf{A}_{{\text{v}}}(\tilde{\bm{\phi}}_{n_\text{F}}) = [ \mathbf{a}_\text{v}(\tilde{\phi}_{1, n_\text{F}}), {\dots}, \mathbf{a}_\text{v}(\tilde{\phi}_{K_\text{v}, n_\text{F}}) ]{\;\in\;}\mathbb{C}^{N_\text{v}{\times}K_\text{v}},
  \end{equation}
  \begin{equation}
    \mathbf{B}(\tilde{\bm{\tau}}_{n_\text{F}}) = [ \mathbf{b}(\tilde{\tau}_{1, n_\text{F}}), {\dots}, \mathbf{b}(\tilde{\tau}_{K_\text{de}, n_\text{F}}) ]{\;\in\;}\mathbb{C}^{N_\text{sc}{\times}K_\text{de}},
  \end{equation}
  \begin{equation}
    \mathbf{C}(\tilde{\bm{\nu}}_{n_\text{F}}) = [ \mathbf{c}(\tilde{\nu}_{1, n_\text{F}}), {\dots}, \mathbf{c}(\tilde{\nu}_{K_\text{do}, n_\text{F}}) ]{\;\in\;}\mathbb{C}^{N_\text{sym}{\times}K_\text{do}},
  \end{equation}
\end{subequations}
where $K_\text{x}$, $\text{x}{\;\in\;}\left\{ \text{h}, \text{v}, \text{de}, \text{do}\right\}$ denote the number of grids, $\tilde{\bm{\chi}}_{n_\text{F}}, \bm{\chi}{\;\in\;}\{ \bm{\theta}, \bm{\phi}, \bm{\tau}, \bm{\nu} \}$ denote the ADD domain grids. 
Following most prior work on dynamic grids, the numbers of angle, delay, and Doppler domain grids are typically set proportional to the numbers of antennas, subcarriers, and OFDM symbols, respectively, thereby offering trade-offs between computational complexity and model flexibility.

\begin{figure}[!t]
  \centering
  \includegraphics[width = \linewidth]{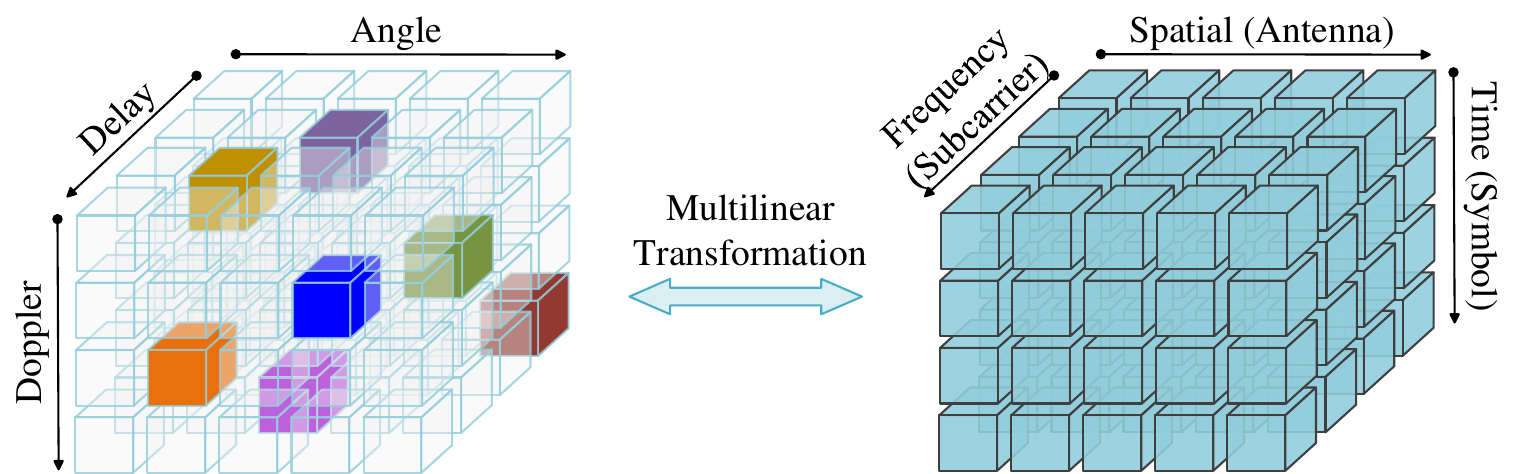}
  \caption{The diagrams of the SFT domain and ADD domain channel tensors. In this case, only the horizontal angle and antenna domains are shown.}
  \label{fig:TensorModelStructuredFactorMatrix}
\end{figure}

Therefore, the SFT domain channels are represented by the Tucker-based model, given by
\begin{equation}\label{eq:multilinear_channel}
    \boldsymbol{\mathcal{H}}_{n_\text{F}} \!=\! \boldsymbol{\mathcal{G}}_{n_\text{F}} {\times_{1}} \mathbf{A}_{\text{h}}(\tilde{\bm{\theta}}_{n_\text{F}}) {\times_{2}} \mathbf{A}_{\text{v}}(\tilde{\bm{\phi}}_{n_\text{F}}) {\times_{3}} \mathbf{B}(\tilde{\bm{\tau}}_{n_\text{F}}) {\times_{4}} \mathbf{C}(\tilde{\bm{\nu}}_{n_\text{F}}),
\end{equation}
where $\boldsymbol{\mathcal{G}}_{n_\text{F}}{\;\in\;}\mathbb{C}^{K_\text{h}{\times}K_\text{v}{\times}K_\text{de}{\times}K_\text{do}}$ denotes ADD domain channels, with elements being the path gain of the specific ADD domain grid and corresponding to the specific rank-one multi-path component.
Based on this representation, channel prediction can be achieved by ADD domain channel acquisition, which leverages the inter-antenna, inter-subcarrier, and inter-symbol correlations.
The specific SFT and ADD domain channels are illustrated in \figref{fig:TensorModelStructuredFactorMatrix}, where only the horizontal angle and antenna domains are shown for visualization. It can be observed that the ADD domain channels contain non-zero values contributed by only a few scatterers, revealing the inherent low-rank structure of SFT domain channels.
\begin{remark}
  The Tucker-based model captures the multi-linear structure of channels, offering the following advantages:
  \begin{itemize}
      \item \textbf{Concise representation}: The high-order structure of the ADD domain channels allows for the concise representation of high-order neighbors in structured prior models.
      \item \textbf{Efficient computation}: The low-dimensional factor matrix in each domain enables significantly reduced computational complexity of the multi-linear model.
  \end{itemize}
\end{remark}

\section{Problem Formulation}\label{sec:problem_formulation}
Although the multi-linear rank is crucial for ADD domain channel acquisition, its optimal selection remains challenging due to the presence of noise.
This challenge motivates the introduction of Bayesian perspectives, which inherently enable automatic rank determination and eliminate the need for the explicit estimation \cite{thomas2020bp, chang2021sparse, cheng2022towards, xu2022sparse}.
Toward this end, we develop probabilistic models for channel prediction with temporal non-stationarity, formulating the dynamic channel prediction problem based on the VFE minimization. Following this, the online approximation and dual-layer optimization of the VFE is achieved through the structure design of trail beliefs.

\subsection{Probabilistic Models}\label{sec:probabilistic_model}
\subsubsection{Observation and Multi-linear Transformation}
Based on the signal model, the PDF of the observation model is expressed as
\begin{equation}\label{eq:observation_model}
    \mathsf{P}(\boldsymbol{\mathcal{Y}}_{(N_\text{F})} \!\mid\! \boldsymbol{\mathcal{H}}_{(N_\text{F})}) = \prod_{n_\text{F} = 1}^{N_\text{F}} \underbrace{\mathsf{CN}(\boldsymbol{\mathcal{Y}}_{n_\text{F}}; \boldsymbol{\mathcal{H}}_{n_\text{F}}, \boldsymbol{\mathcal{C}}({\sigma}_{z}^{2}))}_{\mathsf{P}_{Y, n_\text{F}}},
\end{equation}
where $\boldsymbol{\mathcal{Y}}_{(N_\text{F})}{\;\triangleq\;}\{\boldsymbol{\mathcal{Y}}_{n_\text{F}}\}_{n_\text{F}=1}^{N_\text{F}}$ and $\boldsymbol{\mathcal{H}}_{(N_\text{F})}{\;\triangleq\;}\{\boldsymbol{\mathcal{H}}_{n_\text{F}}\}_{n_\text{F}=1}^{N_\text{F}}$ denote the sets of observations and SFT domain channels for $N_\text{F}$ frames, respectively.

For multi-linear transformations, it is challenging to explicitly capture the dependencies between SFT domain channels and ADD domain grids.
To simplify these dependencies, we uniformly sample the ADD domain to replace the dynamic grids in \eqref{eq:multilinear_channel} with fixed sampling grids. This sampling strategy provides accurate approximations of the continuous physical parameters when the numbers of antennas, subcarriers, and OFDM symbols are sufficiently large.
However, the deviations between sampling grids and ground-truth parameters are inevitable in the practical systems with limited dimensions, which can be captured through ADD domain perturbation parameters on top of uniformly sampling grids.
Specifically, we define the ADD domain perturbed grids as $\tilde{\bm{\chi}}_{n_\text{F}} {\;\triangleq\;} \bar{\bm{\chi}} + \bm{\Delta}\bm{\chi}_{n_\text{F}}$ for $\bm{\chi}{\;\in\;}\{ \bm{\theta}, \bm{\phi}, \bm{\tau}, \bm{\nu} \}$, where $\bar{\bm{\chi}}$ and $\bm{\Delta}\bm{\chi}_{n_\text{F}}$ denote the sampling grids and perturbation parameters, respectively.
Although the resolution of uniform sampling is typically determined by the discrete Fourier transform, slight oversampling can further improve model accuracy with only a marginal increase in the computational complexity.
By treating perturbation parameters as imperfection parameters due to the off-grid effect, the PDF of multi-linear transformation is given as
\begin{subequations}
    \begin{equation}
        \mathsf{P}(\boldsymbol{\mathcal{H}}_{(N_\text{F})} \!\mid\! \boldsymbol{\mathcal{G}}_{(N_\text{F})}; \mathcal{P}_{\text{OG}, (N_\text{F})}) = \prod_{n_\text{F} = 1}^{N_\text{F}} \underbrace{\mathsf{P}(\boldsymbol{\mathcal{H}}_{n_\text{F}} \!\mid\! \boldsymbol{\mathcal{G}}_{n_\text{F}}; \mathcal{P}_{\text{OG}, n_\text{F}})}_{\mathsf{P}_{HG, n_\text{F}}}, 
    \end{equation}
    \begin{align}\label{eq:multi-linear_model}
        \mathsf{P}_{HG, n_\text{F}} {\;\propto\;} 
        {\delta}(\boldsymbol{\mathcal{H}}_{n_\text{F}}& \!-\! \boldsymbol{\mathcal{G}}_{n_\text{F}} {\times_{1}} \mathbf{A}_{\text{h}}(\bar{\bm{\theta}} \!+\! \bm{\Delta}\bm{\theta}_{n_\text{F}}) {\times_{2}} \mathbf{A}_{\text{v}}(\bar{\bm{\phi}} \!+\! \bm{\Delta}\bm{\theta}_{n_\text{F}})  \nonumber \\
        & {\times_{3}} \mathbf{B}(\bar{\bm{\tau}} \!+\! \bm{\Delta}\bm{\tau}_{n_\text{F}}) {\times_{4}} \mathbf{C}(\bar{\bm{\nu}} \!+\! \bm{\Delta}\bm{\nu}_{n_\text{F}}) ), 
    \end{align}
\end{subequations}
where $\boldsymbol{\mathcal{G}}_{(N_\text{F})} {\;\triangleq\;} \{ \boldsymbol{\mathcal{G}}_{n_\text{F}} \}_{n_\text{F}=1}^{N_\text{F}}$ denotes the set of ADD domain channels for $N_\text{F}$ frames, $\mathcal{P}_{\text{OG}, (N_\text{F})}{\;\triangleq\;}\cup_{n_\text{F}=1}^{N_\text{F}} \mathcal{P}_{\text{OG}, n_\text{F}}$ denotes the sets of perturbation parameters in all frames, and $\mathcal{P}_{\text{OG}, n_\text{F}}{\;\triangleq\;}\{ \bm{\Delta}\bm{\theta}_{n_\text{F}}, \bm{\Delta}\bm{\phi}_{n_\text{F}}, \bm{\Delta}\bm{\tau}_{n_\text{F}}, \bm{\Delta}\bm{\nu}_{n_\text{F}} \}$.

\subsubsection{ADD Domain Channels}
Beyond sparsity, the ADD domain channels exhibit long-timescale correlations due to spatial consistency and cross-domain correlations due to the cluster scattering, which can be captured to further enhance channel prediction. Specifically, the ADD domain channels are modeled by two hidden random processes, expressed as
\begin{equation}\label{eq:hidden_process}
  \mathsf{P}(\boldsymbol{\mathcal{G}}_{(N_\text{F})} \!\mid\! \boldsymbol{\mathcal{S}}_{(N_\text{F})}, \boldsymbol{\mathcal{Q}}_{(N_\text{F})}) = \prod_{n_\text{F} = 1}^{N_\text{F}} \underbrace{\delta(\boldsymbol{\mathcal{G}}_{n_\text{F}} - \boldsymbol{\mathcal{S}}_{n_\text{F}} {\;\odot\;} \boldsymbol{\mathcal{Q}}_{n_\text{F}})}_{\mathsf{P}_{GSQ, n_\text{F}}},
\end{equation}
where $\boldsymbol{\mathcal{S}}_{(N_\text{F})} {\;\triangleq\;} \{ \boldsymbol{\mathcal{S}}_{n_\text{F}} \}_{n_\text{F} = 1}^{N_\text{F}}$, and $\boldsymbol{\mathcal{Q}}_{(N_\text{F})} {\;\triangleq\;} \{ \boldsymbol{\mathcal{Q}}_{n_\text{F}} \}_{n_\text{F} = 1}^{N_\text{F}}$, the binary-valued tensor $\boldsymbol{\mathcal{S}}_{n_\text{F}}$ and complex-valued tensor $\boldsymbol{\mathcal{Q}}_{n_\text{F}}$ describe the sparsity and power of the ADD domain channel, respectively. 
Since ADD domain channel tensor elements that are not visible in one frame may become visible in subsequent frames under temporally non-stationary environments, it is crucial to employ hidden value tensors for the modeling of all potential path gains \cite{lian2019exploiting}.

\begin{figure}[!t]
  \centering
  \includegraphics[width = \linewidth]{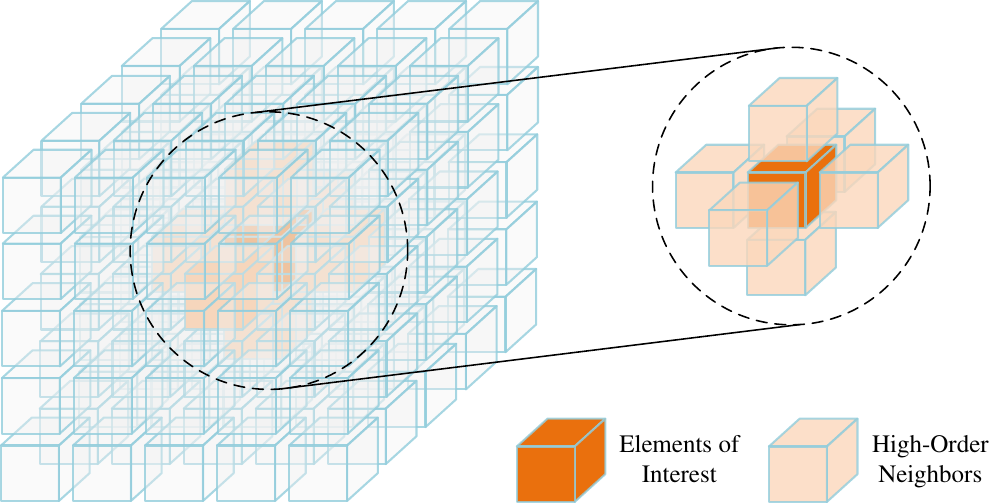}
  \caption{The three-dimensional example of high-order neighbors. In this case, we have $\mathcal{N}{\;\triangleq\;}\{(0, 0, {\pm}1), (0, {\pm}1, 0), ({\pm}1, 0, 0)\}$.}
  \label{fig:Action_Region}
\end{figure}

Due to the smooth variation of the physical parameters of each path across frames, the hidden support tensor is modeled as a first-order Markov process, given by
\begin{equation}
  \mathsf{P}(\boldsymbol{\mathcal{S}}_{(N_\text{F})}; \boldsymbol{\mathcal{M}}) = \prod_{n_\text{F} = 1}^{N_\text{F}} \mathsf{P}(\boldsymbol{\mathcal{S}}_{n_\text{F}} \!\mid\! \boldsymbol{\mathcal{S}}_{n_\text{F} - 1}; \boldsymbol{\mathcal{M}}),
\end{equation}
where $\boldsymbol{\mathcal{S}}_{0} {\;\triangleq\;} \boldsymbol{\mathcal{C}}(0)$, $\boldsymbol{\mathcal{M}}$ denotes the transition factor and controls the sparsity transition probability across frames. In this model, $\mathsf{P}(\boldsymbol{\mathcal{S}}_{n_\text{F}} \!\mid\! \boldsymbol{\mathcal{S}}_{n_\text{F} - 1}; \boldsymbol{\mathcal{M}})$ denotes the transition probabilistic models of the hidden support tensor, given by
\begin{equation}\label{eq:hidden_support_model}
  \mathsf{P}(\boldsymbol{\mathcal{S}}_{n_\text{F}} \!\mid\! \boldsymbol{\mathcal{S}}_{n_\text{F} - 1}; \boldsymbol{\mathcal{M}}) = \mathsf{P}_{\text{MRF}}(\boldsymbol{\mathcal{S}}_{n_\text{F}}) \mathsf{P}_{\text{TC}}(\boldsymbol{\mathcal{S}}_{n_\text{F}} \!\mid\! \boldsymbol{\mathcal{S}}_{n_\text{F} - 1}; \boldsymbol{\mathcal{M}}),
\end{equation}
where $\mathsf{P}_{\text{TC}}(\boldsymbol{\mathcal{S}}_{n_\text{F}} \!\mid\! \boldsymbol{\mathcal{S}}_{n_\text{F} - 1}; \boldsymbol{\mathcal{M}})$ denotes the long-timescale probabilistic model across frames, given by
\begin{align}
  \mathsf{P}_{\text{TC}}&(\boldsymbol{\mathcal{S}}_{n_\text{F}} \!\mid\! \boldsymbol{\mathcal{S}}_{n_\text{F} - 1}; \boldsymbol{\mathcal{M}}) {\;\propto\;} \nonumber \\
  & \mathsf{exp}(\langle \boldsymbol{\mathcal{M}}, (2\boldsymbol{\mathcal{S}}_{n_\text{F} - 1}-\boldsymbol{\mathcal{C}}(1)) {\;\odot\;} (2\boldsymbol{\mathcal{S}}_{n_\text{F}}-\boldsymbol{\mathcal{C}}(1)) \rangle),
\end{align}
where the conversion from binary to bipolar support tensor is inspired by the Ising model \cite{zhang2020variance, xu2023joint, liu2023structured} and $\mathsf{P}_{\text{MRF}}(\boldsymbol{\mathcal{S}}_{n_\text{F}})$ denotes the intra-frame cross-domain probabilistic model, given by
\begin{equation}
    \mathsf{P}_{\text{MRF}}(\boldsymbol{\mathcal{S}}_{n_\text{F}}) {\propto} \prod_{\mathbf{r}{\in}\mathcal{N}} \mathsf{exp}( {\gamma} \langle (2\boldsymbol{\mathcal{S}}_{n_\text{F}} - \boldsymbol{\mathcal{C}}(1)), (2\boldsymbol{\mathcal{S}}_{n_\text{F}}^{[\mathbf{r}]} - \boldsymbol{\mathcal{C}}(1))\rangle ),
\end{equation}
where ${\gamma}$ controls the strength of cross-domain correlations, and $\mathcal{N}$ denotes the relative index set of high-order neighbors, whose three-dimensional example is shown in \figref{fig:Action_Region}.

Similarly, the hidden value tensor is modeled based on the hidden value tensor of the previous frame, given by
\begin{equation}\label{eq:hidden_value_model}
  \boldsymbol{\mathcal{Q}}_{n_\text{F}} = (\boldsymbol{\mathcal{C}}(1) - \boldsymbol{\mathcal{L}}) {\;\odot\;} \boldsymbol{\mathcal{Q}}_{n_\text{F}-1} + \boldsymbol{\mathcal{L}}{\;\odot\;}\boldsymbol{\mathcal{W}}_{n_\text{F}},
\end{equation}
where $\boldsymbol{\mathcal{L}}$ denotes the transition factor and controls the long-timescale temporal correlations, $\boldsymbol{\mathcal{Q}}_{0} {\;\triangleq\;} \boldsymbol{\mathcal{C}}(0)$, $\boldsymbol{\mathcal{W}}_{n_\text{F}}$ following TCGD controls the variations of the hidden value tensor across frames, given by
\begin{equation}
  \mathsf{P}(\boldsymbol{\mathcal{W}}) {\propto} \mathsf{CN}(\boldsymbol{\mathcal{W}}; \boldsymbol{\mathcal{C}}(0), \boldsymbol{\mathcal{V}}) \prod_{\mathbf{r}{\in}\mathcal{N}} \mathsf{CN}(\boldsymbol{\mathcal{W}}; \boldsymbol{\mathcal{C}}(0), {\gamma}^{-1}\boldsymbol{\mathcal{V}}^{[\mathbf{r}]}).
\end{equation}
It implies that the power of each element in $\boldsymbol{\mathcal{W}}_{n_\text{F}}$ is controlled not only by its own hyperparameter but also by the hyperparameters within $\mathcal{N}$.
\begin{remark}
  The temporal variations of hyperparameters are assumed to be negligible over the duration of interest, enabling joint processing of multiple frames to enhance hyperparameter learning. This assumption is typically satisfied by selecting the proper number of frames for hyperparameter learning.
\end{remark}

\subsection{Dual-Layer Online VFE Minimization Formulation}\label{sec:TSDCPFormulation}

\begin{figure}[!t]
  \centering
  \includegraphics[width = \linewidth]{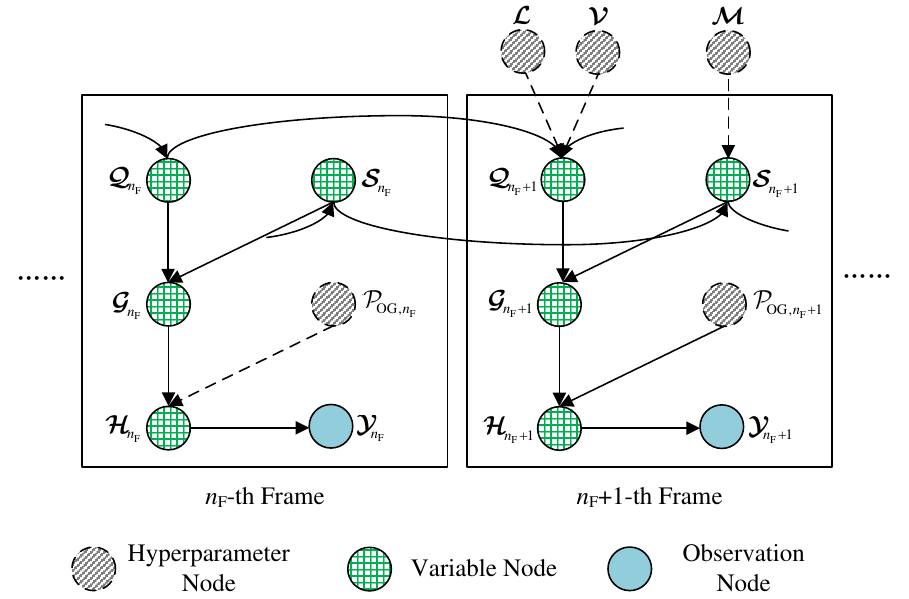}
  \caption{Hierarchical Bayesian model for TS-DCP problem.}
  \label{fig:HierarchicalBayesianModel}
\end{figure}

Building on the above probabilistic models, the hierarchical Bayesian model is depicted in \figref{fig:HierarchicalBayesianModel}, where only the interaction between two frames is shown. Therefore, the acquisition of ADD domain channels based on the minimum mean square error (MMSE) criterion is expressed as
\begin{equation}\label{eq:MMSE_estimator}
  \hat{\boldsymbol{\mathcal{G}}}_{(N_\text{F})} = \int \boldsymbol{\mathcal{G}}_{(N_\text{F})} \mathsf{P}( \boldsymbol{\mathcal{G}}_{(N_\text{F})} \!\mid\! \boldsymbol{\mathcal{Y}}_{(N_\text{F})}; \mathcal{P}_{\text{OG}, (N_\text{F})}, \mathcal{P}_{\text{HP}} ),
\end{equation}
where $\mathsf{P}( \boldsymbol{\mathcal{G}}_{(N_\text{F})} \!\mid\! \boldsymbol{\mathcal{Y}}_{(N_\text{F})}; \mathcal{P}_{\text{OG}, (N_\text{F})}, \mathcal{P}_{\text{HP}} )$ denotes the posterior PDF, $\mathcal{P}_{\text{HP}} {\;\triangleq\;} \{ \boldsymbol{\mathcal{M}}, \boldsymbol{\mathcal{L}}, \boldsymbol{\mathcal{V}} \}$ denotes the set of hyperparameters.

Since the posterior PDFs involve multiple integrals of the joint posterior PDF, the exact MMSE estimator suffers from prohibitive computational complexity in massive MIMO systems. 
To address this challenge, we employ the variational inference approach to approximate the complicated PDF by trial beliefs with simplified structures \cite{zhang2018advances}.
Following the KL-divergence minimization criterion, the joint PDF approximation is given by
\begin{equation}\label{eq:joint_VFE_minimization}
  \hat{\mathsf{b}} = \mathop{\arg\min}\limits_{\mathsf{b}} F_{\mathrm{V}},
\end{equation}
where $\mathsf{b}{\;\triangleq\;}\mathsf{b}( \boldsymbol{\mathcal{H}}_{(N_\text{F})}, \boldsymbol{\mathcal{G}}_{(N_\text{F})}, \boldsymbol{\mathcal{S}}_{(N_\text{F})}, \boldsymbol{\mathcal{Q}}_{(N_\text{F})}; \mathcal{P}_{\text{OG}, (N_\text{F})}, \mathcal{P}_{\text{HP}} )$ denotes the trial belief, $F_{\mathrm{V}}$ denotes the joint VFE and is defined as the KL-divergence between the trial belief and joint PDF. 

Because the variables and hyperparameters are shared across frames, a straightforward VFE minimization involves storing all received signals and jointly optimizing the trial beliefs for all frame. However, this prevents the real-time inference and is meaningless for the channel prediction task. To circumvent this issue, we simplify the trial belief to the factorized form inspired by variational messaging algorithms \cite{winn2005variational}, given as
\begin{align}\label{eq:hyperpara_factor}
  \mathsf{b} ( &\boldsymbol{\mathcal{H}}_{(N_\text{F})},  \boldsymbol{\mathcal{G}}_{(N_\text{F})}, \boldsymbol{\mathcal{S}}_{(N_\text{F})}, \boldsymbol{\mathcal{Q}}_{(N_\text{F})}; \mathcal{P}_{\text{OG}, (N_\text{F})}, \mathcal{P}_{\text{HP}}) {\;\approx\;} \nonumber \\
  & \mathsf{b}(\mathcal{P}_{\text{HP}}) \mathsf{b}(\mathcal{P}_{\text{OG}, (N_\text{F})}) \mathsf{b} ( \boldsymbol{\mathcal{H}}_{(N_\text{F})}, \boldsymbol{\mathcal{G}}_{(N_\text{F})}, \boldsymbol{\mathcal{S}}_{(N_\text{F})}, \boldsymbol{\mathcal{Q}}_{(N_\text{F})}),
\end{align}
where the hyperparameter and perturbation parameter beliefs are constrained to uninformative Dirac-Delta functions parameterized by unknown ground-truth, given as
\begin{subequations}\label{eq:hyperpara_belief_HP}
  \begin{equation}
    \mathsf{b}(\mathcal{P}_{\text{HP}}) = \mathsf{b}(\boldsymbol{\mathcal{M}}) \mathsf{b}(\boldsymbol{\mathcal{L}}) \mathsf{b}(\boldsymbol{\mathcal{V}}),
  \end{equation}
  \begin{equation}
    \mathsf{b}(\boldsymbol{\mathcal{P}}) = \delta(\boldsymbol{\mathcal{P}} - \hat{\boldsymbol{\mathcal{P}}}), \boldsymbol{\mathcal{P}} {\;\in\;} \{ \boldsymbol{\mathcal{M}}, \boldsymbol{\mathcal{L}}, \boldsymbol{\mathcal{V}} \},
  \end{equation}
\end{subequations}
and
\begin{subequations}\label{eq:hyperpara_belief_ML}
  \begin{equation}
    \mathsf{b}(\mathcal{P}_{\text{OG}, (N_\text{F})}) = \prod_{n_\text{F}=1}^{N_\text{F}} \mathsf{b}(\mathcal{P}_{\text{OG}, n_\text{F}}),
  \end{equation}
  \begin{equation}
      \mathsf{b}(\mathcal{P}_{\text{OG}, n_\text{F}}) =  \mathsf{b}(\bm{\Delta}\bm{\theta}_{n_\text{F}}) \mathsf{b}(\bm{\Delta}\bm{\phi}_{n_\text{F}}) \mathsf{b}(\bm{\Delta}\bm{\tau}_{n_\text{F}}) \mathsf{b}(\bm{\Delta}\bm{\nu}_{n_\text{F}}),
  \end{equation}
  \begin{equation}
    \mathsf{b}(\bm{\chi}) = {\delta}( \bm{\chi} - \hat{\bm{\chi}} ), \bm{\chi}{\;\in\;}\{ \bm{\Delta}\bm{\theta}_{n_\text{F}}, \bm{\Delta}\bm{\phi}_{n_\text{F}}, \bm{\Delta}\bm{\tau}_{n_\text{F}}, \bm{\Delta}\bm{\nu}_{n_\text{F}} \},
  \end{equation}
\end{subequations}
where $\hat{\boldsymbol{\mathcal{P}}}$ and $\hat{\bm{\chi}}$ denote the ground-truth hyperparameters and perturbation parameters, respectively. This simplification is justified by the observation that the model hyperparameters $\mathcal{P}_{\text{HP}}$, which capture long-timescale and cross-domain correlations in ADD domain channels, vary much more slowly than the instantaneous channels. Therefore, the model hyperparameters are assumed to remain stable within each frame, as is the case for perturbation parameters. For the shared variables across successive frames, we approximate them by the posterior mean \cite{liu2021sparse} and ignore the posterior variance. This further decouples the trial belief into the factorized form of per-frame trial belief, given by
\begin{align}\label{eq:belief_online_approx}
  \mathsf{b} ( \boldsymbol{\mathcal{H}}_{(N_\text{F})}, \boldsymbol{\mathcal{G}}_{(N_\text{F})}, & \boldsymbol{\mathcal{S}}_{(N_\text{F})}, \boldsymbol{\mathcal{Q}}_{(N_\text{F})}) {\;\approx\;} \nonumber\\
  &\prod_{n_\text{F}=1}^{N_\text{F}} \mathsf{b}( \boldsymbol{\mathcal{H}}_{n_\text{F}}, \boldsymbol{\mathcal{G}}_{n_\text{F}}, \boldsymbol{\mathcal{S}}_{n_\text{F}}, \boldsymbol{\mathcal{Q}}_{n_\text{F}}),
\end{align}
where $\mathsf{b}( \boldsymbol{\mathcal{H}}_{n_\text{F}}, \boldsymbol{\mathcal{G}}_{n_\text{F}}, \boldsymbol{\mathcal{S}}_{n_\text{F}}, \boldsymbol{\mathcal{Q}}_{n_\text{F}})$ denotes the trial belief of the $n_\text{F}$-th frame.

Building on the structure design of trial beliefs, the joint VFE minimization is reformulated as the dual-layer optimization. The outer-layer focuses on perturbation parameter and hyperparameter learning, while the inner-layer aims to minimize the VFE with learned parameters from the outer-layer. 
Specifically, the inner-layer operates in the online manner, restricting the optimization of trial beliefs to a single frame, as described in the following proposition.
\begin{proposition}\label{prop:online_approx}
    Given belief structures in \eqref{eq:hyperpara_factor}, \eqref{eq:hyperpara_belief_HP}, \eqref{eq:hyperpara_belief_ML}, annd \eqref{eq:belief_online_approx}, the VFE optimization in the inner-layer is given by
    \begin{equation}\label{eq:onlineVFEM}
      \hat{\mathsf{b}}_{n_\text{F}} = \mathop{\arg\min}\limits_{ \mathsf{b}_{n_\text{F}}} \bar{F}_{\mathrm{V}, n_\text{F}},
    \end{equation}
    where $\mathsf{b}_{n_\text{F}}{\;\triangleq\;}\mathsf{b}( \boldsymbol{\mathcal{H}}_{n_\text{F}}, \boldsymbol{\mathcal{G}}_{n_\text{F}}, \boldsymbol{\mathcal{S}}_{n_\text{F}}, \boldsymbol{\mathcal{Q}}_{n_\text{F}})$ denotes the per-frame trial belief, $\bar{F}_{\mathrm{V}, n_\text{F}}$ denotes the per-frame VFE given by \eqref{eq:per_frame_VFE_hyperparameter_approx} at the top of the next page, $\hat{\mathcal{P}}_{\text{HP}}{\;\triangleq\;}\{ \hat{\boldsymbol{\mathcal{M}}}, \hat{\boldsymbol{\mathcal{L}}}, \hat{\boldsymbol{\mathcal{V}}} \}$ and $\hat{\mathcal{P}}_{\text{OG}, n_\text{F}}{\;\triangleq\;}\{ \hat{\bm{\Delta}}\bm{\theta}_{n_\text{F}}, \hat{\bm{\Delta}}\bm{\phi}_{n_\text{F}}, \hat{\bm{\Delta}}\bm{\tau}_{n_\text{F}}, \hat{\bm{\Delta}}\bm{\nu}_{n_\text{F}} \}$ denote the set of hyperparameters and perturbation parameters learned from the outer-layer, respectively.
    \begin{proof}
        Please refer to \appref{app:online_approx}.
    \end{proof}
\end{proposition}

\begin{figure*}[!t]
  \normalsize
  \begin{equation}\label{eq:per_frame_VFE_hyperparameter_approx}
    \bar{F}_{\text{V}, n_\text{F}} {\;\triangleq\;} \mathsf{D}[ \mathsf{b}( \boldsymbol{\mathcal{H}}_{n_\text{F}}, \boldsymbol{\mathcal{G}}_{n_\text{F}}, \boldsymbol{\mathcal{S}}_{n_\text{F}}, \boldsymbol{\mathcal{Q}}_{n_\text{F}}) \parallel \mathsf{P} (\boldsymbol{\mathcal{H}}_{n_\text{F}}, \boldsymbol{\mathcal{G}}_{n_\text{F}}, \boldsymbol{\mathcal{S}}_{n_\text{F}}, \boldsymbol{\mathcal{Q}}_{n_\text{F}}, \boldsymbol{\mathcal{Y}}_{n_\text{F}} \mid \boldsymbol{\mathcal{S}}_{n_\text{F} - 1}, \boldsymbol{\mathcal{Q}}_{n_\text{F} - 1}; \hat{\mathcal{P}}_{\text{OG}, n_\text{F}}, \hat{\mathcal{P}}_{\text{HP}} ) ],
  \end{equation}

  \hrulefill
  \vspace*{4pt}
\end{figure*}

\section{Online Tensor-Structured Dynamic Channel Prediciton Algorithm}\label{sec:online_TSDCP}

Building upon the dual-layer VFE optimization process, the inference of the inner- and outer-layers is discussed in \secref{sec:inner_layer} and \secref{sec:outer_layer}, respectively, and ultimately yields the online TS-DCP algorithm.

\subsection{Multi-Linear Inference with Structured Priors}\label{sec:inner_layer}

\begin{figure}[!t]
  \centering
  \includegraphics[width = \linewidth]{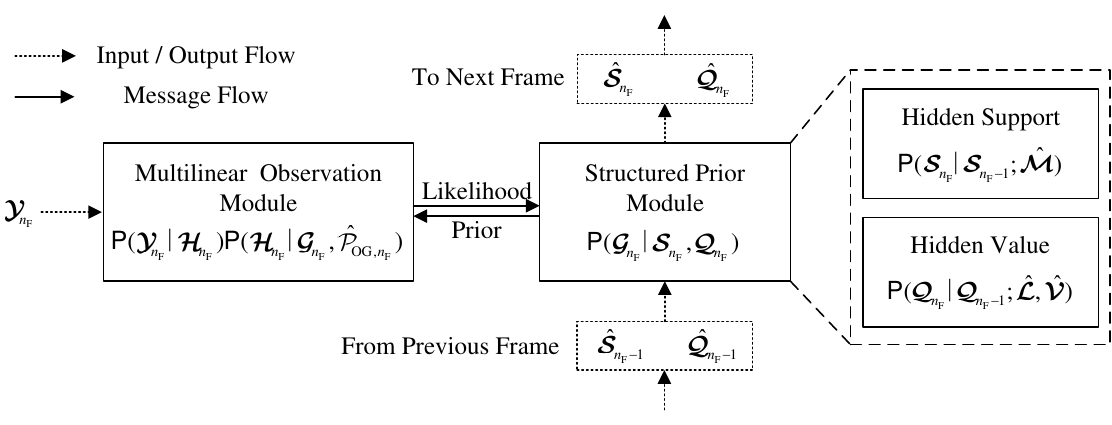}
  \caption{Module diagram of the multi-linear inference with structured priors.}
  \label{fig:DynamicChannelTrackingModule}
\end{figure}

The VFE minimization in \eqref{eq:onlineVFEM} is formulated as the alternating minimization between two modules, as shown in \figref{fig:DynamicChannelTrackingModule}. The first module, the multi-linear observation (MO) module, focuses on observation and multi-linear transformation models, while the second, the structured prior (SP) module, incorporates ADD domain channel models. Therefore, the trial belief of the $n_\text{F}$-th frame is constrained by
\begin{equation}
    \mathsf{b}_{n_\text{F}} = \mathsf{b}_{\text{MO}, n_\text{F}}  \mathsf{b}_{\text{HS}, n_\text{F}}  \mathsf{b}_{\text{HV}, n_\text{F}},
\end{equation}
where the $\mathsf{b}_{\text{MO}, n_\text{F}}$ is optimized in the MO module, while $\mathsf{b}_{\text{HS}, n_\text{F}}$ and $\mathsf{b}_{\text{HV}, n_\text{F}}$ are optimized in the SP module.

\subsubsection{MO Module}
Since the MO model is a multi-linear generalization of the linear observation model, we employ the Bethe method \cite{yedidia2005constructing, zhang2021unifying} based on joint PDF factorization and region partitioning and constrain the trial belief as 
\begin{equation}\label{eq:structured_trial_belief_multi-linear_module}
  \mathsf{b}_{\text{MO}, n_\text{F}} = \frac{\mathsf{b}_{Y, n_\text{F}} \mathsf{b}_{HG, n_\text{F}} \mathsf{b}_{G, n_\text{F}} }{\mathsf{f}_{H, n_\text{F}} \mathsf{f}_{G, n_\text{F}}^{N}}
\end{equation}
where $\mathsf{b}_{Y, n_\text{F}}$, $\mathsf{b}_{HG, n_\text{F}}$, and $\mathsf{b}_{G, n_\text{F}}$ denote the factor beliefs for the observation, multi-linear transformation, and prior models, respectively, $\mathsf{f}_{H, n_\text{F}}$ and $\mathsf{f}_{G, n_\text{F}}$ denote the variable beliefs for $\boldsymbol{\mathcal{H}}_{n_\text{F}}$ and $\boldsymbol{\mathcal{G}}_{n_\text{F}}$, respectively, and $N {\;\triangleq\;} N_\text{h}N_\text{v}N_\text{sc}N_\text{sym}$ denotes the repetitive counts to be excluded in the Bethe method.

Therefore, the VFE in the MO module is converted to the Bethe free energy (BFE), expressed as
\begin{align}\label{eq:BFE_MO}
  F_{\text{MO}, n_\text{F}} = &\mathsf{D}[ \mathsf{b}_{Y, n_\text{F}} \parallel \mathsf{P}_{Y, n_\text{F}} ] + \mathsf{D}[\mathsf{b}_{HG, n_\text{F}} \parallel \mathsf{P}_{HG, n_\text{F}}] + \nonumber \\
  & \mathsf{D}[\mathsf{b}_{G, n_\text{F}} \parallel \mathsf{P}_{G, n_\text{F}}^{\text{pri}} ] + \mathsf{H}[ \mathsf{f}_{H, n_\text{F}} ] + N\mathsf{H}[ \mathsf{f}_{G, n_\text{F}} ],
\end{align}
where $\mathsf{P}_{G, n_\text{F}}^{\text{pri}}$ denotes the prior PDF approximation of ADD domain channels from the SP module. 
To guarantee global dependencies between beliefs, we impose marginal consistency constraints (MCCs) for factor and variable beliefs sharing the same variables. However, sinde the MCCs for continuous random variables remain intractable, we relax them to first-order (FO) and second-order (SO) sufficient statistical consistency constraints (SSCCs), denoted as
\begin{subequations}\label{eq:MO_SSCC}
  \begin{equation}
    \mathsf{E}_{\mathsf{b}_{Y, n_\text{F}}}\{ \boldsymbol{\mathcal{H}}_{n_\text{F}}\} = \mathsf{E}_{\mathsf{b}_{HG, n_\text{F}}}\{ \boldsymbol{\mathcal{H}}_{n_\text{F}} \} = \mathsf{E}_{\mathsf{f}_{H, n_\text{F}}}\{ \boldsymbol{\mathcal{H}}_{n_\text{F}} \},
  \end{equation}
  \begin{equation}
    \mathsf{E}_{\mathsf{b}_{HG, n_\text{F}}}\{ \boldsymbol{\mathcal{G}}_{n_\text{F}} \} = \mathsf{E}_{\mathsf{b}_{G, n_\text{F}}}\{ \boldsymbol{\mathcal{G}}_{n_\text{F}} \} = \mathsf{E}_{\mathsf{f}_{G, n_\text{F}}}\{ \boldsymbol{\mathcal{G}}_{n_\text{F}} \},
  \end{equation}
  \begin{equation}
    \mathsf{V}_{\mathsf{b}_{Y, n_\text{F}}}\{ \boldsymbol{\mathcal{H}}_{n_\text{F}}\} = \mathsf{V}_{\mathsf{b}_{HG, n_\text{F}}}\{ \boldsymbol{\mathcal{H}}_{n_\text{F}} \} = \mathsf{V}_{\mathsf{f}_{H, n_\text{F}}}\{ \boldsymbol{\mathcal{H}}_{n_\text{F}} \},
  \end{equation}
  \begin{equation}
    \mathsf{V}_{\mathsf{b}_{HG, n_\text{F}}}\{ \boldsymbol{\mathcal{G}}_{n_\text{F}} \} = \mathsf{V}_{\mathsf{b}_{G, n_\text{F}}}\{ \boldsymbol{\mathcal{G}}_{n_\text{F}} \} = \mathsf{V}_{\mathsf{f}_{G, n_\text{F}}}\{ \boldsymbol{\mathcal{G}}_{n_\text{F}} \}.
  \end{equation}
\end{subequations}
Building upon the above BFE construction and MCC relaxation, the inference in the MO module is formulated as the constrained BFE minimization problem, given by
\begin{equation}\label{eq:MO_CBFEM}
  \mathop{\arg\min}\limits_{\mathsf{b}_{\text{MO}, n_\text{F}}} F_{\text{MO}, n_\text{F}}, {\quad} \text{s.t.} \eqref{eq:MO_SSCC},
\end{equation}
whose fixed-point equations are summarized in \propref{prop:MO_FP}.
\begin{proposition}\label{prop:MO_FP}
    Based on the Lagrange multiplier method \cite{boyd2004convex}, the fixed-point equations of \eqref{eq:MO_CBFEM} are given by
    \begin{subequations}
        \begin{equation}\label{eq:fixed_point_b_Y}
            \mathsf{b}_{Y, n_\text{F}} {\;\propto\;} \mathsf{P}_{Y, n_\text{F}} \mathsf{CN}(\boldsymbol{\mathcal{H}}_{n_\text{F}}; \boldsymbol{\mathcal{N}}_{n_\text{F}}^{H,b_{H}}, -(\boldsymbol{\mathcal{E}}_{n_\text{F}}^{H,b_{H}})^{{\odot}-1}),
        \end{equation}
        \begin{align}\label{eq:fixed_point_b_HG}
            \mathsf{b}_{HG, n_\text{F}} {\;\propto\;} \mathsf{P}_{HG, n_\text{F}} & \mathsf{CN}(\boldsymbol{\mathcal{H}}_{n_\text{F}}; \boldsymbol{\mathcal{N}}_{n_\text{F}}^{H,b_{HG}}, -(\boldsymbol{\mathcal{E}}_{n_\text{F}}^{H,b_{HG}})^{{\odot}-1}) \nonumber \\
            \mathsf{CN}(& \boldsymbol{\mathcal{G}}_{n_\text{F}}; \boldsymbol{\mathcal{N}}_{n_\text{F}}^{G,b_{HG}}, -(\boldsymbol{\mathcal{E}}_{n_\text{F}}^{G,b_{HG}})^{{\odot}-1}),
        \end{align}
        \begin{equation}\label{eq:fixed_point_b_G}
            \mathsf{b}_{G, n_\text{F}} {\;\propto\;} \mathsf{P}_{G, n_\text{F}}^{\text{pri}} \mathsf{CN}(\boldsymbol{\mathcal{G}}_{n_\text{F}}; \boldsymbol{\mathcal{N}}_{n_\text{F}}^{G,b_{G}}, -(\boldsymbol{\mathcal{E}}_{n_\text{F}}^{G,b_{G}})^{{\odot}-1}),
        \end{equation}
        \begin{align}\label{eq:fixed_point_f_H}
            \mathsf{f}_{H, n_\text{F}} {\;\propto\;} & \mathsf{CN}(\boldsymbol{\mathcal{H}}_{n_\text{F}}; \boldsymbol{\mathcal{N}}_{n_\text{F}}^{H,b_{H}}, -(\boldsymbol{\mathcal{E}}_{n_\text{F}}^{H,b_{H}})^{{\odot}-1}) \nonumber \\
            & \mathsf{CN}(\boldsymbol{\mathcal{H}}_{n_\text{F}}; \boldsymbol{\mathcal{N}}_{n_\text{F}}^{H,b_{HG}}, -(\boldsymbol{\mathcal{E}}_{n_\text{F}}^{H,b_{HG}})^{{\odot}-1}),
        \end{align}
        \begin{align}\label{eq:fixed_point_f_G}
            \mathsf{f}_{G, n_\text{F}} {\;\propto\;} & \mathsf{CN}(\boldsymbol{\mathcal{G}}_{n_\text{F}}; \boldsymbol{\mathcal{N}}_{n_\text{F}}^{G,b_{HG}}, -N(\boldsymbol{\mathcal{E}}_{n_\text{F}}^{G,b_{HG}})^{{\odot}-1}) \nonumber \\
            & \mathsf{CN}(\boldsymbol{\mathcal{G}}_{n_\text{F}}; \boldsymbol{\mathcal{N}}_{n_\text{F}}^{G,b_{G}}, -N(\boldsymbol{\mathcal{E}}_{n_\text{F}}^{G,b_{G}})^{{\odot}-1}),
        \end{align}
    \end{subequations}
    where $\boldsymbol{\mathcal{N}}_{n_\text{F}}^{\text{x},b_{\text{x}}}$ is defined in \eqref{eq:fixed_point_variable_definition} at the top of the next page, $\boldsymbol{\mathcal{U}}_{n_\text{F}}^{\text{x}, b_{\text{x}}}$ and  $\boldsymbol{\mathcal{E}}_{n_\text{F}}^{\text{x}, b_{\text{x}}}$ are the Lagrange multipliers for SSCCs.
    \begin{proof}
        Please refer to \appref{app:MO_FP}.
    \end{proof}
\end{proposition}
By substituting fixed-point equations in \propref{prop:MO_FP} into SSCCs, the iteration equations for the Lagrange multipliers can be derived. These equations are organized based on the message flow backward and forward from observations to ADD domain channels, summarized in \alref{alg:MO_VI}.

\begin{figure*}[!t]
  \normalsize
  \begin{subequations}\label{eq:fixed_point_variable_definition}
        \begin{equation}
            \boldsymbol{\mathcal{N}}_{n_\text{F}}^{H,b_{H}} = -\boldsymbol{\mathcal{U}}_{n_\text{F}}^{H,b_{H}} {\;\oslash\;} \boldsymbol{\mathcal{E}}_{n_\text{F}}^{H,b_{H}} + \mathsf{E}_{\mathsf{b}_{Y, n_\text{F}}}\{ \boldsymbol{\mathcal{H}}_{n_\text{F}}\}, \boldsymbol{\mathcal{N}}_{n_\text{F}}^{H,b_{HG}} = -\boldsymbol{\mathcal{U}}_{n_\text{F}}^{H,b_{HG}} {\;\oslash\;} \boldsymbol{\mathcal{E}}_{n_\text{F}}^{H,b_{HG}} + \mathsf{E}_{\mathsf{b}_{Y, n_\text{F}}}\{ \boldsymbol{\mathcal{H}}_{n_\text{F}}\}
        \end{equation}
        \begin{equation}
            \boldsymbol{\mathcal{N}}_{n_\text{F}}^{G,b_{HG}} = -\boldsymbol{\mathcal{U}}_{n_\text{F}}^{G,b_{HG}} {\;\oslash\;} \boldsymbol{\mathcal{E}}_{n_\text{F}}^{G,b_{HG}} + \mathsf{E}_{\mathsf{b}_{G, n_\text{F}}}\{ \boldsymbol{\mathcal{G}}_{n_\text{F}}\}, \boldsymbol{\mathcal{N}}_{n_\text{F}}^{G,b_{G}} = -\boldsymbol{\mathcal{U}}_{n_\text{F}}^{G,b_{G}} {\;\oslash\;} \boldsymbol{\mathcal{E}}_{n_\text{F}}^{G,b_{G}} + \mathsf{E}_{\mathsf{b}_{G, n_\text{F}}}\{ \boldsymbol{\mathcal{G}}_{n_\text{F}}\}
        \end{equation}
    \end{subequations}
  \hrulefill
\end{figure*}

\begin{algorithm}[!t]
  \caption{Single Iteration of MO Module}
  \label{alg:MO_VI}
  \begin{algorithmic}[1]
  \Require {}
  \Statex {$\boldsymbol{\mathcal{Y}}_{n_\text{F}}$: Observation tensor.}
  \Statex {$\mathsf{P}_{G, n_\text{F}}^{\text{pri}}$: Message of $\boldsymbol{\mathcal{G}}_{n_\text{F}}$ from SP module.}
  \Statex {$\hat{\mathcal{P}}_{\text{OG}, n_\text{F}}$: Estimation result of ${\mathcal{P}}_{\text{OG}, n_\text{F}}$.}
  \Statex {${\sigma}_{z}^{2}$: Noise variance}  
  \Ensure {}
  \Statex {$\mathsf{P}_{G, n_\text{F}}^\text{lik}$: Message of $\boldsymbol{\mathcal{G}}_{n_\text{F}}$ to SP module.}
  \Statex {$\hat{\boldsymbol{\mathcal{G}}}_{n_\text{F}}$: Estimation result of ${\boldsymbol{\mathcal{G}}}_{n_\text{F}}$.}
  \State {$\boldsymbol{\mathcal{E}}_{n_\text{F}}^{G, b_{HG}} = -(\mathsf{V}_{\mathsf{b}_{G, n_\text{F}}}\{ \boldsymbol{\mathcal{G}}_{n_\text{F}}\})^{{\odot}-2} - \boldsymbol{\mathcal{E}}_{n_\text{F}}^{G, b_{G}} / N$} \label{stage:OM_start}
  \State {$\begin{aligned}
      \boldsymbol{\mathcal{E}}_{n_\text{F}}^{H, b_{H}} & = ( (\boldsymbol{\mathcal{E}}_{n_\text{F}}^{G, b_{HG}})^{{\odot}-1} {\;\times_{1}\;} |\mathbf{A}_{\text{h}}(\hat{\bm{\theta}}_{n_\text{F}}) |^{{\odot}2} \nonumber \\
      {\;\times_{2}\;} & |\mathbf{A}_{\text{v}}(\hat{\bm{\phi}}_{n_\text{F}}) |^{{\odot}2} {\;\times_{3}\;} |\mathbf{B}(\hat{\bm{\tau}}_{n_\text{F}})|^{{\odot}2} {\;\times_{4}\;} |\mathbf{C}(\hat{\bm{\nu}}_{n_\text{F}})|^{{\odot}2} )^{{\odot}-1}
  \end{aligned}$}\label{step:state_MO_1}
  \State {$\begin{aligned}
      \boldsymbol{\mathcal{N}}_{n_\text{F}}^{H, b_{H}} & = -\boldsymbol{\mathcal{U}}_{n_\text{F}}^{H, b_{H}} {\;\oslash\;} \boldsymbol{\mathcal{E}}_{n_\text{F}}^{H, b_{H}} + \hat{\boldsymbol{\mathcal{G}}}_{n_\text{F}}{\;\times_{1}\;} \mathbf{A}_{\text{h}}(\hat{\bm{\theta}}_{n_\text{F}}) \nonumber \\
      {\;\times_{2}\;} & \mathbf{A}_{\text{v}}(\hat{\bm{\phi}}_{n_\text{F}}) {\;\times_{3}\;} \mathbf{B}(\hat{\bm{\tau}}_{n_\text{F}}) {\;\times_{4}\;} \mathbf{C}(\hat{\bm{\nu}}_{n_\text{F}})
  \end{aligned}$}\label{step:state_MO_2}
  \State {$\mathsf{P}_{H, n_\text{F}}^{\text{pri}} = \mathsf{CN}( \boldsymbol{\mathcal{H}}_{n_\text{F}}; \boldsymbol{\mathcal{N}}_{n_\text{F}}^{H, b_{H}}, -(\boldsymbol{\mathcal{E}}_{n_\text{F}}^{H, b_{H}})^{{\odot}-1} ) $}
  \State {$\mathsf{b}_{H, n_\text{F}} {\;\propto\;} \mathsf{P}_{Y, n_\text{F}} \mathsf{P}_{H, n_\text{F}}^{\text{pri}} $, $\hat{\boldsymbol{\mathcal{H}}}_{n_\text{F}} = \mathsf{E}_{\mathsf{b}_{H, n_\text{F}}}\{ \boldsymbol{\mathcal{H}}_{n_\text{F}} \}$}
  \State {$\boldsymbol{\mathcal{E}}_{n_\text{F}}^{H, b_{HG}} = -(\mathsf{V}_{\mathsf{b}_{H, n_\text{F}}}\{ \boldsymbol{\mathcal{H}}_{n_\text{F}} \})^{{\odot}-2} - \boldsymbol{\mathcal{E}}_{n_\text{F}}^{H, b_{H}}$}
  \State {$\bm{\Delta}\boldsymbol{\mathcal{N}}_{n_\text{F}} = \boldsymbol{\mathcal{U}}_{n_\text{F}}^{H, b_{H}} {\;\oslash\;} \boldsymbol{\mathcal{E}}_{n_\text{F}}^{H, b_{HG}} + \hat{\boldsymbol{\mathcal{H}}}_{n_\text{F}} - \boldsymbol{\mathcal{N}}_{n_\text{F}}^{H, b_{H}}$}
  \State {$\boldsymbol{\mathcal{U}}_{n_\text{F}}^{H, b_{H}} = \bm{\Delta}\boldsymbol{\mathcal{N}}_{n_\text{F}} {\;\oslash\;} ((\boldsymbol{\mathcal{E}}_{n_\text{F}}^{H, b_{H}})^{{\odot}-1} + (\boldsymbol{\mathcal{E}}_{n_\text{F}}^{H, b_{HG}})^{{\odot}-1})$}
  \State {$\boldsymbol{\mathcal{J}}_{n_\text{F}}^{H, b_{H}} = \boldsymbol{\mathcal{E}}_{n_\text{F}}^{H, b_{H}} {\;\odot\;} ( 1 + \boldsymbol{\mathcal{E}}_{n_\text{F}}^{H, b_{H}} {\;\odot\;} (\mathsf{V}_{\mathsf{b}_{H, n_\text{F}}}\{ \boldsymbol{\mathcal{H}}_{n_\text{F}} \})^{{\odot}2} )$}
  \State {$\begin{aligned}
      \boldsymbol{\mathcal{J}}_{n_\text{F}}^{G, b_{G}} & = \boldsymbol{\mathcal{J}}_{n_\text{F}}^{H, b_{H}} {\;\times_{1}\;} | \mathbf{A}_{\text{h}}^{H}(\hat{\bm{\theta}}_{n_\text{F}})|^{{\odot}2} {\;\times_{2}\;} | \mathbf{A}_{\text{v}}^{H}(\hat{\bm{\phi}}_{n_\text{F}})|^{{\odot}2} \nonumber \\
      {\;\times_{3}\;} &|\mathbf{B}^{H}(\hat{\bm{\tau}}_{n_\text{F}})|^{{\odot}2} {\;\times_{4}\;} |\mathbf{C}^{H}(\hat{\bm{\nu}}_{n_\text{F}})|^{{\odot}2}
  \end{aligned}$}\label{step:state_MO_3}
  \State {$\boldsymbol{\mathcal{E}}_{n_\text{F}}^{G, b_{G}} = ( (\boldsymbol{\mathcal{J}}_{n_\text{F}}^{G, b_{G}})^{{\odot}-1} - (N\boldsymbol{\mathcal{E}}_{n_\text{F}}^{H, b_{HG}})^{{\odot}-1} )^{{\odot}-1}$}
  \State {$\begin{aligned}
      \boldsymbol{\mathcal{N}}_{n_\text{F}}^{G, b_{G}} &= \hat{\boldsymbol{\mathcal{G}}}_{n_\text{F}} + ( \boldsymbol{\mathcal{E}}_{n_\text{F}}^{G, b_{G}} )^{{\odot}-1} {\;\odot\;} (\boldsymbol{\mathcal{U}}_{n_\text{F}}^{R, b_{R}}  {\;\times_{1}\;} \hat{\mathbf{A}}_{\text{h}}^{H}(\hat{\bm{\theta}}_{n_\text{F}}) \nonumber \\
      {\;\times_{2}\;} &\hat{\mathbf{A}}_{\text{v}}^{H}(\hat{\bm{\phi}}_{n_\text{F}}) {\;\times_{3}\;} \hat{\mathbf{B}}^{H}(\hat{\bm{\tau}}_{n_\text{F}}) {\;\times_{4}\;} \hat{\mathbf{C}}^{H}(\hat{\bm{\nu}}_{n_\text{F}}) )
  \end{aligned}$}\label{step:state_MO_4}
  \State {$\mathsf{P}_{G, n_\text{F}}^\text{lik} = \mathsf{CN}(\boldsymbol{\mathcal{G}}_{n_\text{F}}; \boldsymbol{\mathcal{N}}_{n_\text{F}}^{G, b_{G}}, -(\boldsymbol{\mathcal{E}}_{n_\text{F}}^{G, b_{G}})^{{\odot}-1} )$}
  \State {$\mathsf{b}_{G, n_\text{F}}{\;\propto\;}\mathsf{P}_{G, n_\text{F}}^\text{pri} \mathsf{P}_{G, n_\text{F}}^\text{lik} $, $\hat{\boldsymbol{\mathcal{G}}}_{n_\text{F}} = \mathsf{E}_{\mathsf{b}_{G, n_\text{F}}}\{ \boldsymbol{\mathcal{G}}_{n_\text{F}} \}$}
  \end{algorithmic}
\end{algorithm}  

\subsubsection{SP Module}
In this module, due to correlations between elements in hidden support tensors, the structured trial beliefs are still constructed by the Bethe method, given by
\begin{equation}
  \mathsf{b}_{\text{HS}, n_\text{F}} = \frac{\mathsf{b}_{S, n_\text{F}}^{\text{lik}} \mathsf{b}_{S, n_\text{F}}^{\text{MRF}} \mathsf{b}_{S, n_\text{F}}^{\text{TC}} }{\mathsf{f}_{S, n_\text{F}}^{F+1}} ,
\end{equation}
where $\mathsf{b}_{S, n_\text{F}}^{\text{lik}}$, $\mathsf{b}_{S, n_\text{F}}^{\text{MRF}}$, and $\mathsf{b}_{S, n_\text{F}}^{\text{TC}}$ denote the factor beliefs introduced for likelihood PDF, MRF model, and long-timescale correlation model, respectively, $\mathsf{f}_{S, n_\text{F}}$ denotes the variable belief introduced for $\boldsymbol{\mathcal{S}}_{n_\text{F}}$, $F$ denotes the number of high-order neighbors. Therefore, the VFE of hidden support tensors in the SP module is expressed as
\begin{align}\label{eq:BFE_HS}
  F_{\text{HS}, n_\text{F}} = & \mathsf{D}[ \mathsf{b}_{S, n_\text{F}}^{\text{lik}} \parallel \mathsf{P}_{S, n_\text{F}}^{\text{lik}}] + \mathsf{D}[\mathsf{b}_{S, n_\text{F}}^{\text{MRF}} \parallel \mathsf{P}_{S, n_\text{F}}^{\text{MRF}} ] + \nonumber \\  
  \mathsf{D} & [\mathsf{b}_{S, n_\text{F}}^{\text{TC}} \parallel \mathsf{P}_{S, n_\text{F}}^{\text{TC}, \text{OL}}] + (F + 1)\mathsf{H}[ \mathsf{f}_{S, n_\text{F}} ] ,
\end{align}
where $\mathsf{P}_{S, n_\text{F}}^{\text{lik}}$ denotes the likelihood PDF of hidden support tensors, and $\mathsf{P}_{S, n_\text{F}}^{\text{TC}, \text{OL}}$ denotes the temporal correlation model with an online approximation, defined by $\mathsf{P}_{S, n_\text{F}}^{\text{TC}, \text{OL}}{\;\triangleq\;}\mathsf{P}_{\text{TC}}(\boldsymbol{\mathcal{S}}_{n_\text{F}} \!\mid\! \hat{\boldsymbol{\mathcal{S}}}_{n_\text{F} - 1}; \hat{\boldsymbol{\mathcal{M}}})$.
Furthermore, since the elements in the hidden support tensor are binary-valued, the MCCs are equivalent to FO-SSCCs, given by
\begin{equation}\label{eq:HS_MCC}
  \mathsf{E}_{\mathsf{b}_{S, n_\text{F}}^{\text{MRF}}}\{ \boldsymbol{\mathcal{S}}_{n_\text{F}} \}\!=\!\mathsf{E}_{\mathsf{b}_{S, n_\text{F}}^{\text{TC}}}\{ \boldsymbol{\mathcal{S}}_{n_\text{F}} \}\!=\!\mathsf{E}_{\mathsf{b}_{S, n_\text{F}}^{\text{lik}}}\{ \boldsymbol{\mathcal{S}}_{n_\text{F}} \}\!=\!\mathsf{E}_{\mathsf{f}_{S, n_\text{F}}}\{ \boldsymbol{\mathcal{S}}_{n_\text{F}} \},
\end{equation}
where $\mathsf{b}_{S, n_\text{F}}^{\text{MRF}}$ denotes the factor belief of MRFs, given by
\begin{equation}
  \mathsf{b}_{S, n_\text{F}}^{\text{MRF}}(\boldsymbol{\mathcal{S}}_{n_\text{F}}) = \prod_{\mathbf{r}{\in}\mathcal{N}} \mathsf{b}_{S, n_\text{F}, \mathbf{r}}(\boldsymbol{\mathcal{S}}_{n_\text{F}}, \boldsymbol{\mathcal{S}}_{n_\text{F}}^{[\mathbf{r}]}).
\end{equation}
Therefore, the inference of hidden support tensors in the SP module is formulated as
\begin{equation}\label{eq:HS_CBFEM}
  \mathop{\arg\min}\limits_{\mathsf{b}_{\text{HS}, n_\text{F}}} F_{\text{HS}, n_\text{F}}, {\quad} \text{s.t.} \eqref{eq:HS_MCC},
\end{equation}
and its fixed-point equations can be organized as the belief propagation rules.

\begin{algorithm}[!t]
  \caption{Single Iteration of SP Module}
  \label{alg:SP_VI}
  \begin{algorithmic}[1]
  \Require {}
  \Statex {$\mathsf{P}_{S, n_\text{F}}^{\text{lik}}$: Message of $\boldsymbol{\mathcal{S}}_{n_\text{F}}$ from MO module.}
  \Statex {$\mathsf{P}_{Q, n_\text{F}}^{\text{lik}}$: Message of $\boldsymbol{\mathcal{Q}}_{n_\text{F}}$ from MO module.}
  \Statex {$\hat{\mathcal{P}}_{\text{HP}}$: Estimation result of ${\mathcal{P}}_{\text{HP}}$.}
  \Ensure {}
  \Statex {$\mathsf{P}_{S, n_\text{F}}^{\text{pri}}$: Message of $\boldsymbol{\mathcal{S}}_{n_\text{F}}$ to MO module.}
  \Statex {$\mathsf{P}_{Q, n_\text{F}}^{\text{pri}}$: Message of $\boldsymbol{\mathcal{Q}}_{n_\text{F}}$ to MO module.}
  \State {$\bar{\boldsymbol{\mathcal{U}}}_{n_\text{F}, \mathbf{r}}^{\text{MRF}} = \sum_{\bar{\mathbf{r}}{\in}\mathcal{N}{\backslash}\mathbf{r}}\boldsymbol{\mathcal{U}}_{n_\text{F}, \bar{\mathbf{r}}}^{\text{MRF}}, \forall \mathbf{r} {\;\in\;} \mathcal{N}$}
  \State {$\boldsymbol{\mathcal{U}}_{n_\text{F}, \mathbf{r}}^{S, \text{MRF}} = \boldsymbol{\mathcal{U}}_{n_\text{F}}^{\text{lik}} + \boldsymbol{\mathcal{U}}_{n_\text{F}}^{\text{TC}} + \bar{\boldsymbol{\mathcal{U}}}_{n_\text{F}, \mathbf{r}}^{\text{MRF}}, \forall \mathbf{r} {\;\in\;} \mathcal{N}$}
  \State {$\boldsymbol{\mathcal{U}}_{n_\text{F}, \mathbf{r}}^{\text{MRF}, 1} = 1 + \mathsf{exp}(2\boldsymbol{\mathcal{C}}({\gamma}) + \boldsymbol{\mathcal{U}}_{n_\text{F}, \mathbf{r}}^{S, \text{MRF}}), \forall \mathbf{r} {\;\in\;} \mathcal{N}$}
  \State {$\boldsymbol{\mathcal{U}}_{n_\text{F}, \mathbf{r}}^{\text{MRF}, 2} = \mathsf{exp}(2\boldsymbol{\mathcal{C}}({\gamma})) + \mathsf{exp}(\boldsymbol{\mathcal{U}}_{n_\text{F}, \mathbf{r}}^{S, \text{MRF}}), \forall \mathbf{r} {\;\in\;} \mathcal{N}$}
  \State {$\boldsymbol{\mathcal{U}}_{n_\text{F}, \mathbf{r}}^{\text{MRF}} = \boldsymbol{\mathcal{U}}_{n_\text{F}, \mathbf{r}}^{\text{MRF}, 1} - \boldsymbol{\mathcal{U}}_{n_\text{F}, \mathbf{r}}^{\text{MRF}, 2}, \forall \mathbf{r} {\;\in\;} \mathcal{N}$}
  \State {$\boldsymbol{\mathcal{U}}_{n_\text{F}}^{\text{MRF}} = \sum_{\mathbf{r}{\in}\mathcal{N}}\boldsymbol{\mathcal{U}}_{n_\text{F}, \mathbf{r}}^{\text{MRF}}, \forall \mathbf{r} {\;\in\;} \mathcal{N}$}
  \State {$\mathsf{P}_{S, n_\text{F}}^{\text{pri}} {\;\propto\;} \mathsf{exp}(-\langle \boldsymbol{\mathcal{U}}_{n_\text{F}}^{\text{TC}} + \boldsymbol{\mathcal{U}}_{n_\text{F}}^{\text{MRF}}, \boldsymbol{\mathcal{S}}_{n_\text{F}} \rangle)$}
  \State {$\mathsf{b}_{S, n_\text{F}} {\;\propto\;} \mathsf{P}_{S, n_\text{F}}^{\text{lik}} \mathsf{P}_{S, n_\text{F}}^{\text{pri}}, \mathsf{b}_{Q, n_\text{F}} {\;\propto\;} \mathsf{P}_{Q, n_\text{F}}^{\text{lik}} \mathsf{P}_{Q, n_\text{F}}^{\text{pri}} $}
  \State {$\hat{\boldsymbol{\mathcal{S}}}_{n_\text{F}} = \mathsf{E}_{\mathsf{b}_{S, n_\text{F}}}\{ \boldsymbol{\mathcal{S}}_{n_\text{F}} \}$, $\hat{\boldsymbol{\mathcal{Q}}}_{n_\text{F}} = \mathsf{E}_{\mathsf{b}_{Q, n_\text{F}}}\{ \boldsymbol{\mathcal{Q}}_{n_\text{F}} \}$}
  \end{algorithmic}
\end{algorithm}

Due to the conditional independency of the elements in $\boldsymbol{\mathcal{Q}}_{n_\text{F}}$ given $\hat{\boldsymbol{\mathcal{Q}}}_{n_\text{F}-1}$, no additional constraints are required on the trial beliefs of hidden value tensors, thus simplifying the VFE minimization. By directly minimizing the KL-divergence, the fixed-point equations of hidden value tensors in the SP module are expressed as 
\begin{equation}\label{eq:HV_FP}
    \mathsf{b}_{\text{HV}, n_\text{F}} {\;\propto\;} \mathsf{P}_{Q, n_\text{F}}^{\text{lik}} \mathsf{P}_{Q, n_\text{F}}^{\text{TC},\text{OL}},
\end{equation}
respectively, where $\mathsf{P}_{Q, n_\text{F}}^{\text{lik}}$ and $\mathsf{P}_{Q, n_\text{F}}^{\text{TC},\text{OL}}$ denote the likelihood PDF and the temporal correlation model with online approximation of hidden value tensors, respectively, and we define $\mathsf{P}_{Q, n_\text{F}}^{\text{TC},\text{OL}} {\;\triangleq\;} \mathsf{P}(\boldsymbol{\mathcal{Q}}_{n_\text{F}} \mid \hat{\boldsymbol{\mathcal{Q}}}_{n_\text{F} - 1}; \hat{\boldsymbol{\mathcal{L}}}, \hat{\boldsymbol{\mathcal{V}}})$.

Similar to the MO module, the iterative equations of Lagrange multipliers in the SP module are summarized in \alref{alg:SP_VI}. In this context, $\boldsymbol{\mathcal{U}}_{n_\text{F}}^{\text{lik}}$ and $\boldsymbol{\mathcal{U}}_{n_\text{F}}^{\text{TC}}$ denote log-likelihood ratios of $\mathsf{P}_{S, n_\text{F}}^{\text{lik}}$ and $\mathsf{P}_{Q, n_\text{F}}^{\text{lik}}$, respectively.

\subsubsection{Closing the Loops}
Based on the iterative equations within the MO and SP modules, a critical step in closing the loop of online TS-DCP algorithm is the message exchange between these modules, including the prior PDFs of the ADD domain channel $\mathsf{P}_{G, n_\text{F}}^{\text{pri}}$, the likelihood PDFs of the hidden support tensor $\mathsf{P}_{S, n_\text{F}}^{\text{lik}}$, and the likelihood PDFs of the hidden values tensor $\mathsf{P}_{Q, n_\text{F}}^{\text{lik}}$.
Based on the joint PDF factorization and the stationary point of beliefs, the joint posterior PDF of $\boldsymbol{\mathcal{G}}_{n_\text{F}}$, $\boldsymbol{\mathcal{S}}_{n_\text{F}}$, and $\boldsymbol{\mathcal{Q}}_{n_\text{F}}$ is 
\begin{equation}
    \mathsf{P}( \boldsymbol{\mathcal{G}}_{n_\text{F}}, \boldsymbol{\mathcal{S}}_{n_\text{F}}, \boldsymbol{\mathcal{Q}}_{n_\text{F}} \mid \boldsymbol{\mathcal{Y}}_{n_\text{F}} ) {\;\propto\;}  \mathsf{P}_{GSQ, n_\text{F}} \mathsf{P}_{S, n_\text{F}}^{\text{pri}} \mathsf{P}_{Q, n_\text{F}}^{\text{pri}} \mathsf{P}_{G, n_\text{F}}^{\text{lik}}.
\end{equation}
Therefore, the prior and likelihood PDFs can be calculated by marginalizing the joint posterior PDF and excluding the contributions of the input data. However, the zero elements in $\boldsymbol{\mathcal{S}}_{n_\text{F}}$ result in the unobservability of the corresponding elements in $\boldsymbol{\mathcal{Q}}_{n\text{F}}$, thus yielding the uninformative likelihood PDFs and preventing the VFE minimization.
To mitigate this issue, we adopt the threshold-based Gaussian sum approximation \cite{ziniel2013dynamic} in the likelihood PDF of $\boldsymbol{\mathcal{Q}}_{n_\text{F}}$.

\subsection{Perturbation Parameter and Hyperparameter Learning}\label{sec:outer_layer}
Beyond the MO and SP modules, the perturbation parameter and hyperparameter learning refinies the grids and hyperparameters for MO and SP modules, respectively.

\subsubsection{Perturbation Parameter}
To reduce computational complexity, we employ the sequential optimization approach for the perturbation parameter.
In the example of the horizontal angle domain, the perturbation parameter learning rule based on the VFE minimization is formulated as
\begin{align}\label{eq:horizontal_angle_VFEM}
  \hat{\bm{\Delta}}\bm{\theta}_{n_\text{F}}^{\star} &\mathop{=}^{(a)} \mathop{\arg\min}_{\hat{\bm{\Delta}}\bm{\theta}_{n_\text{F}}} \mathsf{D}[\mathsf{b}_{HG, n_\text{F}} \parallel \mathsf{P}(\boldsymbol{\mathcal{H}}_{n_\text{F}} \!\mid\! \boldsymbol{\mathcal{G}}_{n_\text{F}}; \hat{\mathcal{P}}_{\text{OG}, n_\text{F}})] \nonumber \\
  &= \mathop{\arg\max}_{\hat{\bm{\Delta}}\bm{\theta}_{n_\text{F}}} \underbrace{\int \mathsf{b}_{HG, n_\text{F}} \ln \mathsf{P}(\boldsymbol{\mathcal{H}}_{n_\text{F}}\!\mid\!\boldsymbol{\mathcal{G}}_{n_\text{F}}; \hat{\mathcal{P}}_{\backslash\text{h}, n_\text{F}})}_{{\;\triangleq\;}f(\hat{\bm{\Delta}}\bm{\theta}_{n_\text{F}})},
\end{align}
where $(a)$ is due to only this KL divergence term depending on $\hat{\bm{\Delta}}\bm{\theta}_{n_\text{F}}$, $\hat{\mathcal{P}}_{{\backslash}\text{h}, n_\text{F}} {\;\triangleq\;} \{\hat{\bm{\Delta}}\bm{\theta}_{n_\text{F}}, \hat{\bm{\Delta}}\bm{\phi}_{n_\text{F}}^{\star}, \hat{\bm{\Delta}}\bm{\tau}_{n_\text{F}}^{\star}, \hat{\bm{\Delta}}\bm{\nu}_{n_\text{F}}^{\star} \}$ includes the horizontal angle domain perturbation parameters currently being optimized and the previously optimized perturbation parameters of other domains. Through the first-order Taylor series of the factor matrix with respect to the perturbation parameters, the objective function can be approximated as the quadratic function, shown in \propref{prop:perturbation}.
\begin{proposition}\label{prop:perturbation}
    $f(\hat{\bm{\Delta}}{\theta}_{n_\text{F}})$ can be approximated as the quadratic function with respect to $\hat{\bm{\Delta}}{\theta}_{n_\text{F}}$, given by
    \begin{equation}
        f(\hat{\bm{\Delta}}{\theta}_{n_\text{F}}) = \hat{\bm{\Delta}}\bm{\theta}_{n_\text{F}}^{T} \bm{\Pi}_{\text{h}} \hat{\bm{\Delta}}\bm{\theta}_{n_\text{F}} - 2\bm{\mu}_{\text{h}}^{T}\hat{\bm{\Delta}}\bm{\theta}_{n_\text{F}} + C,
    \end{equation}
    where $C$ denotes the constant term independent of $\hat{\bm{\Delta}}{\theta}_{n_\text{F}}$, $\bm{\Pi}_{\text{h}}$ and $\bm{\mu}_{\text{h}}$ are given by
    \begin{subequations}
        \begin{align}
            \bm{\Pi}_{\text{h}} = (\dot{\mathbf{A}}_{\text{h}}^{H} &(\bar{\bm{\theta}})\dot{\mathbf{A}}_{\text{h}}(\bar{\bm{\theta}}))^{\ast} {\odot} \nonumber \\
            & \sum_{n} (\hat{\mathbf{g}}_{n_\text{F}, n}\hat{\mathbf{g}}_{n_\text{F}, n}^{H} + \mathsf{diag}\{ \bm{\varepsilon}_{G, n_\text{F}, n}^{\text{h}} \}),
        \end{align}
        \begin{align}
            \bm{\mu}_{\text{h}} = & \sum_{n} \mathsf{Re}\{ \mathsf{diag}^{H}\{ \hat{\mathbf{g}}_{n_\text{F}, n}^{\text{h}} \} \dot{\mathbf{A}}_{\text{h}}^{H}(\bar{\bm{\theta}}) {\Delta}\hat{\mathbf{h}}_{n_\text{F}, n}^{(1)} \} - \nonumber \\
            &\sum_{n} \mathsf{Re}\{ \mathsf{diag}\{ \dot{\mathbf{A}}_{\text{h}}^{H}(\bar{\bm{\theta}}) \mathbf{A}_{\text{h}}(\bar{\bm{\theta}})\} {\;\odot\;} \bm{\varepsilon}_{G, n_\text{F}, n}^{\text{h}} \},
        \end{align}
    \end{subequations}
    where $\dot{\mathbf{A}}_{\text{h}}(\bm{\theta})$ denotes the first-order derivative matrix of $\mathbf{A}_{\text{h}}(\bm{\theta})$ with respect to $\bm{\theta}$, $\mathbf{g}_{n_\text{F}, n}^{\text{h}}$, $\bm{\varepsilon}_{G, n_\text{F}, n}^{\text{h}}$, and ${\Delta}\hat{\mathbf{h}}_{n_\text{F}, n}^{(1)}$ denote the fibers of $\hat{\boldsymbol{\mathcal{G}}}_{n_\text{F}}^{\text{h}}$, $\boldsymbol{\mathcal{E}}_{G, n_\text{F}}^{\text{h}}$, and ${\Delta}\hat{\boldsymbol{\mathcal{H}}}_{n_\text{F}}$ with respect to the first dimension, respectively, $\hat{\boldsymbol{\mathcal{G}}}_{n_\text{F}}^{\text{h}}$, $\boldsymbol{\mathcal{E}}_{G, n_\text{F}}^{\text{h}}$, and ${\Delta}\hat{\boldsymbol{\mathcal{H}}}_{n_\text{F}}$ are defined by \eqref{eq:auxiliary_tensor} at the top of the next page. 
    \begin{proof}
        Please refer to \appref{app:perturbation}.
    \end{proof}
\end{proposition}
Therefore, the perturbation parameter learning is formulated as the quadratic optimization problem, which can provide the optimal solution by well-known approaches with low-complexity, as are the learning rules in other domains.
By leveraging efficient tensor operations, the auxiliary matrices and vectors in the quadratic optimization problems can also be constructed in the low-complexity manner.

\begin{figure*}[!t]
  \normalsize
  \begin{subequations}\label{eq:auxiliary_tensor}
    \begin{equation}
        \hat{\boldsymbol{\mathcal{G}}}_{n_\text{F}}^{\text{h}} = \hat{\boldsymbol{\mathcal{G}}}_{n_\text{F}}{\;\times_{2}\;} \mathbf{A}_{\text{v}}(\bar{\bm{\phi}} + \hat{\bm{\Delta}}\bm{\phi}_{n_\text{F}}^{\star}) {\;\times_{3}\;} \mathbf{B}(\bar{\bm{\tau}} + \hat{\bm{\Delta}}\bm{\tau}_{n_\text{F}}^{\star}) {\;\times_{4}\;} \mathbf{C}(\bar{\bm{\nu}} + \hat{\bm{\Delta}}\bm{\nu}_{n_\text{F}}^{\star}),
    \end{equation}
    \begin{equation}
        \boldsymbol{\mathcal{E}}_{G, n_\text{F}}^{\text{h}} {\triangleq} \boldsymbol{\mathcal{E}}_{G, n_\text{F}}{\times_{2}} |\mathbf{A}_{\text{v}}(\bar{\bm{\phi}} + \hat{\bm{\Delta}}\bm{\phi}_{n_\text{F}}^{\star})|^{{\odot}2} {\times_{3}} |\mathbf{B}(\bar{\bm{\tau}} + \hat{\bm{\Delta}}\bm{\tau}_{n_\text{F}}^{\star})|^{{\odot}2} {\times_{4}} |\mathbf{C}(\bar{\bm{\nu}} + \hat{\bm{\Delta}}\bm{\nu}_{n_\text{F}}^{\star})|^{{\odot}2},
    \end{equation}
    \begin{equation}
        {\Delta}\hat{\boldsymbol{\mathcal{H}}}_{n_\text{F}} = \hat{\boldsymbol{\mathcal{H}}}_{n_\text{F}} - \hat{\boldsymbol{\mathcal{G}}}_{n_\text{F}} {\;\times_{1}\;} \mathbf{A}_{\text{h}}(\bar{\bm{\theta}}) {\;\times_{2}\;} \mathbf{A}_{\text{v}}(\bar{\bm{\phi}} + \hat{\bm{\Delta}}\bm{\phi}_{n_\text{F}}^{\star}) {\;\times_{3}\;} \mathbf{B}(\bar{\bm{\tau}} + \hat{\bm{\Delta}}\bm{\tau}_{n_\text{F}}^{\star}) {\;\times_{4}\;} \mathbf{C}(\bar{\bm{\nu}} + \hat{\bm{\Delta}}\bm{\nu}_{n_\text{F}}^{\star}).
    \end{equation}
  \end{subequations}
  \hrulefill
\end{figure*}

\subsubsection{Hyperparameter Learning}
Different from the perturbation parameters, the hyperparameters in $\hat{\mathcal{P}}_{\text{HP}}$ are shared across all frames and learned adaptively based on the current observations.
For the ${N}_{\text{F}}$-th frame, the hyperparameter learning rules based on the VFE minimization are expressed as
\begin{subequations}
  \begin{equation}\label{eq:learning_rule_M}
    \hat{\boldsymbol{\mathcal{M}}}^{\star} = \mathop{\arg\max}_{\hat{\boldsymbol{\mathcal{M}}}} \sum_{n_\text{F} = 1}^{{N}_{\text{F}}} \int \mathsf{b}_{S, n_\text{F}} \ln \mathsf{P}_{\text{TC}}(\boldsymbol{\mathcal{S}}_{n_\text{F}} \!\mid\! \boldsymbol{\mathcal{S}}_{n_\text{F} - 1}; \hat{\boldsymbol{\mathcal{M}}}),
  \end{equation}
  \begin{equation}\label{eq:learning_rule_LV}
    \{ \hat{\boldsymbol{\mathcal{L}}}^{\star}, \hat{\boldsymbol{\mathcal{V}}}^{\star} \}  = \mathop{\arg\max}_{\{\hat{\boldsymbol{\mathcal{L}}}, \hat{\boldsymbol{\mathcal{V}}}\}} \sum_{n_\text{F} = 1}^{{N}_{\text{F}}} \int \mathsf{b}_{Q, n_\text{F}} \ln \mathsf{P}(\boldsymbol{\mathcal{Q}}_{n_\text{F}} \!\mid\! \boldsymbol{\mathcal{Q}}_{n_\text{F} - 1}; \hat{\boldsymbol{\mathcal{L}}}, \hat{\boldsymbol{\mathcal{V}}}).
  \end{equation}
\end{subequations}
where other terms in \eqref{eq:onlineVFEM} are independent of $ \hat{\boldsymbol{\mathcal{M}}}$, $\hat{\boldsymbol{\mathcal{L}}}^{\star}$, and $\hat{\boldsymbol{\mathcal{V}}}^{\star}$.
Due to the hyperparameter coupling in hidden value tensors, it is intractable to find an analytic solution for $\hat{\boldsymbol{\mathcal{V}}}^{\star}$.
To address this issue, we first learn hyperparameter $\bar{\boldsymbol{\mathcal{V}}}$ instead and then approximate $\hat{\boldsymbol{\mathcal{V}}}^{\star}$ as \cite{fang2014pattern, fang2016two}
\begin{equation}\label{eq:equivalent_V}
  \hat{\boldsymbol{\mathcal{V}}}^{\star} {\;\approx\;} \bar{\boldsymbol{\mathcal{V}}} + {\gamma} \sum_{\mathbf{r}{\in}\mathcal{N}} \bar{\boldsymbol{\mathcal{V}}}^{[\mathbf{r}]},
\end{equation}
where $\bar{\boldsymbol{\mathcal{V}}}$ is given in \eqref{eq:equivalent_pv_hp} at the top of the next page. Besides, the learning rules of transition and temporal correlation factors are summarized in \propref{prop:hyperparameter}, henceforth all hyperparameters can be learned in the element-wise manner.
\begin{figure*}[!t]
  \normalsize
  \begin{equation}\label{eq:equivalent_pv_hp}
    {\bar{\boldsymbol{\mathcal{V}}}} = \frac{1}{{N}_{\text{F}}}( |\hat{\boldsymbol{\mathcal{Q}}}_{1}|^{{\odot}2} + \boldsymbol{\mathcal{E}}_{Q, 1} + \sum_{n_\text{F}=2}^{{N}_{\text{F}}} (|\hat{\boldsymbol{\mathcal{Q}}}_{n_\text{F}}|^{{\odot}2} + \boldsymbol{\mathcal{E}}_{Q, n_\text{F}} - 2(1-\hat{\boldsymbol{\mathcal{L}}}^{\star}){\;\odot\;} \mathsf{Re}\{ \hat{\boldsymbol{\mathcal{Q}}}_{n_\text{F}-1}^{\ast} {\;\odot\;} \hat{\boldsymbol{\mathcal{Q}}}_{n_\text{F}} \}  + (1-\hat{\boldsymbol{\mathcal{L}}}^{\star}){\;\odot\;} |\hat{\boldsymbol{\mathcal{Q}}}_{n_\text{F}-1}|^{{\odot}2}){\;\oslash\;} (\hat{\boldsymbol{\mathcal{L}}}^{\star})^{{\odot}2} ).
  \end{equation}
  \hrulefill
\end{figure*}
\begin{proposition}\label{prop:hyperparameter}
    The learning rules of transition and temporal correlation factors are given by
    \begin{subequations}
        \begin{equation}
          \hat{\boldsymbol{\mathcal{M}}}^{\star} = \mathsf{ln}(1 + \boldsymbol{\mathcal{K}}_{M}) - \mathsf{ln}(1 - \boldsymbol{\mathcal{K}}_{M}),
        \end{equation}
        \begin{equation}
          ({N}_{\text{F}} - 1)\bar{\boldsymbol{\mathcal{V}}} {\;\odot\;}|\hat{\boldsymbol{\mathcal{L}}}^{\star}|^{{\odot}2} + \boldsymbol{\mathcal{K}}_{L, 1}{\;\odot\;}\hat{\boldsymbol{\mathcal{L}}}^{\star} + \boldsymbol{\mathcal{K}}_{L, 0} = 0, 
        \end{equation}
    \end{subequations}
    where $\boldsymbol{\mathcal{K}}_{M}$, $\boldsymbol{\mathcal{K}}_{L, 1}$ and $\boldsymbol{\mathcal{K}}_{L, 2}$ are defined by
    \begin{subequations}
        \begin{equation}
          \boldsymbol{\mathcal{K}}_{M} = \frac{1}{{N}_{\text{F}}}\sum_{n_\text{F}=1}^{{N}_{\text{F}}} (2\hat{\boldsymbol{\mathcal{S}}}_{n_\text{F}}-1){\;\odot\;}(2\hat{\boldsymbol{\mathcal{S}}}_{n_\text{F}-1}-1),
        \end{equation}
        \begin{equation}
          \boldsymbol{\mathcal{K}}_{L, 1} {\;\triangleq\;} \sum_{n_\text{F}=2}^{{N}_{\text{F}}} ( | \hat{\boldsymbol{\mathcal{Q}}}_{n_\text{F}-1} |^{{\odot}2} - \mathsf{Re}\{ \hat{\boldsymbol{\mathcal{Q}}}_{n_\text{F}-1}^{\ast} {\;\odot\;} \hat{\boldsymbol{\mathcal{Q}}}_{n_\text{F}} \} ), 
        \end{equation}
        \begin{equation}
          \boldsymbol{\mathcal{K}}_{L, 0} {\;\triangleq\;} -\sum_{n_\text{F}=2}^{{N}_{\text{F}}} ( |\hat{\boldsymbol{\mathcal{Q}}}_{n_\text{F}} - \hat{\boldsymbol{\mathcal{Q}}}_{n_\text{F}-1}|^{{\odot}2} + \boldsymbol{\mathcal{E}}_{Q, n_\text{F}} ).
        \end{equation}
    \end{subequations}
    \begin{proof}
        Please refer to \appref{app:hyperparameter}.
    \end{proof}
\end{proposition}

\subsection{Online TS-DCP Algorithm}
Following the dual-layer optimization framework in \secref{sec:TSDCPFormulation}, the proposed online TS-DCP algorithm is summarized as \alref{alg:Dynamic_CT}, where the inner- and outer-layers are given in \secref{sec:inner_layer} and \secref{sec:outer_layer}, respectively. 
Given the estimation of ADD domain channels and perturbation parameters, channel prediction is achieved by the transformation from Doppler domain to temporal domain, expressed as
\begin{align}\label{eq:channel_prediction}
    \boldsymbol{\mathcal{H}}_{n_\text{F}}^{\text{CP}}= \hat{\boldsymbol{\mathcal{G}}}_{n_\text{F}} & {\;\times_{1}\;} \mathbf{A}_{\text{h}}(\bar{\bm{\theta}} + \hat{\bm{\Delta}}{\bm{\theta}}_{n_\text{F}}) {\;\times_{2}\;} \mathbf{A}_{\text{v}}(\bar{\bm{\phi}} + \hat{\bm{\Delta}}\bm{\phi}_{n_\text{F}}) \nonumber \\
    &{\;\times_{3}\;} \mathbf{B}(\bar{\bm{\tau}} + \hat{\bm{\Delta}}{\bm{\tau}}_{n_\text{F}}) {\;\times_{4}\;} \tilde{\mathbf{C}}(\bar{\bm{\nu}} + \hat{\bm{\Delta}}{\bm{\nu}}_{n_\text{F}}), 
\end{align}
where $\tilde{\mathbf{C}}(\cdot)$ denotes the factor matrix in temporal domain for channel prediction. Since \alref{alg:Dynamic_CT} operates in the sliding frame manner, it enables dynamic frame reorganization as new pilot OFDM symbols become available, facilitating real-time channel prediction.
\begin{remark}
    By designing various belief structures and MCC relaxations, we integrate the message passing rules-including variational message passing \cite{winn2005variational}, belief propagation \cite{yedidia2005constructing}, expectation propagation variant \cite{zhang2021unifying}, and expectation maximization \cite{moon1996expectation} into \alref{alg:Dynamic_CT} under the VFE minimization framework. This integration enables different message-passing rule scheduling strategies from the alternating optimization perspective, further reducing computational complexity \cite{hou2024beam}. Moreover, the observation and multi-linear transformation models are compatible with quantized received signals and hybrid analog-digital architectures, offering a practical approach to reducing memory overhead.
\end{remark}

\begin{algorithm}[!t]
\caption{Online TS-DCP Algorithm}
\label{alg:Dynamic_CT}
\begin{algorithmic}[1]
\Require {}
\Statex {$\boldsymbol{\mathcal{Y}}_{(N_\text{F})}$: Observation tensor.}
\Statex {${\sigma}_{z}^{2}$: Noise variance}  
\Ensure {}
\Statex {$\hat{\boldsymbol{\mathcal{G}}}_{(N_\text{F})}$: Estimation result of ${\boldsymbol{\mathcal{G}}}_{(N_\text{F})}$.}
\Statex {$\hat{\mathcal{P}}_{\text{OG}, (N_\text{F})}$: Estimation result of ${\mathcal{P}}_{\text{OG}, (N_\text{F})}$.}
\State {Initialize $\hat{\boldsymbol{\mathcal{M}}}$, $\hat{\boldsymbol{\mathcal{L}}}$, $\hat{\boldsymbol{\mathcal{V}}}$, $\hat{\boldsymbol{\mathcal{S}}}_{0}$, and $\hat{\boldsymbol{\mathcal{Q}}}_{0}$.}
\For {$n_\text{F} = 1, {\dots}, N_\text{F}$}
  \For {$t = 1 : T$}
  \State {Obtain the prior PDF of $\boldsymbol{\mathcal{G}}_{n_\text{F}}$.}
  \State {Execute \alref{alg:MO_VI}.}
  \State {Obtain the likelihood PDFs of $\boldsymbol{\mathcal{S}}_{n_\text{F}}$ and $\boldsymbol{\mathcal{Q}}_{n_\text{F}}$.}
  \State {Execute \alref{alg:SP_VI}.}
  \State {Perturbation parameter / hyperparameter learning.}
  \EndFor
  \State {Predict the channel based on \eqref{eq:channel_prediction}.}
\EndFor
\end{algorithmic}
\end{algorithm}

\section{Simulation Results}\label{sec:simulation_result}
\subsection{Simulation Configuration}
\subsubsection{Scenario Setting}
To validate the exactness of proposed channel and probabilistic models, we employ the QuaDRiGa channel simulator, which generates massive MIMO-OFDM channels consistent with the third Generation Partnership Program (3GPP) New Radio (NR) specifications \cite{3gpp38901} and has been validated in various field trials \cite{jaeckel2014quadriga}. In the QuaDRiGa channel simulator, we consider the 3GPP urban macro (UMa) non-line-of-sight (NLOS) scenarios, where each channel contains $20$ clusters consisting of $20$ subpaths with similar physical parameters.
As MTs traverse the trajectory, the contribution of scatterers evolves with time, and the spatial consistency of the channels is maintained.
Unless specified otherwise, the simulation configurations follow the system model in \secref{sec:system_model}, with parameters summarized in \tabref{tab:simulation_configuration}. 

\begin{table}[!t]
  \caption{Scenario Paramters}
  \label{tab:simulation_configuration}
  \centering
  \begin{tabular}{cc}
  \toprule
  \textbf{Paramter} & \textbf{Value} \\
  \midrule
  Carrier Frequency & $f_{\text{c}} = 6.7$ GHz \\
  Pilot Symbol Interval & ${\Delta}\bar{T} = 14{\;\times\;}35.68$ $\mu$s \\
  Pilot Subcarrier Spacing & ${\Delta}\bar{f} = 4 {\;\times\;} 30$ kHz \\
  Number of BS Antennas & $(N_{\text{h}}, N_\text{v}) = (32, 16)$ \\
  Number of Pilot Subcarriers & $N_\text{sc} = 64$ \\
  Number of Pilot Symbols & $ N_\text{sym} = 8 $ \\
  Height of BS & $h_\text{BS} = 25$ m \\
  Height of MTs & $h_\text{MT} = 1.5$ m \\
  MT Distribution Radius & $r_{\text{MT}} = 200$ m \\
  \bottomrule
  \end{tabular}
\end{table}

\subsubsection{Benchmarks and Performance Metric}
To demonstrate the superiority of the proposed online TS-DCP algorithm, we select the following state-of-the-art algorithms as benchmarks:
\begin{itemize}
  \item \textbf{VKF} \cite{kim2020massive}: Estimates spatial domain channels by the least square (LS) algorithm, with temporal correlations captured by the AR-based vector KF.
  \item \textbf{FIT} \cite{peng2019downlink}: Estimates angle-delay domain channels by the alternating LS (ALS) algorithm, with temporal correlations captured by the first-order Taylor series.
  \item \textbf{PAD} \cite{yin2020addressing}, \textbf{MPAD} \cite{qin2022partial}: Estimates spatial domain channels by linear MMSE (LMMSE) algorithm, with temporal correlations captured by Prony and matrix pencil methods for effective angle-delay domain taps.
\end{itemize}
Besides, we include \textbf{Online TS-DCP} algorithm with pre-sampled grids and unstructured independent prior as benchmarks, referred to as \textbf{Online TS-DCP (PG)} and \textbf{Online TS-DCP (UIP)}, respectively. For benchmarks that only predict channels for future pilot symbols, we predict channels on non-pilot symbols through MMSE interpolation, with prior knowledge of the maximum Doppler frequency and transmission power.
The time-averaged normalized mean square error (TNMSE) of SFT domain channels is adopted as the performance metric, defined by
\begin{equation}
    \text{TNMSE} = \frac{1}{N_\text{F}} \sum_{n_{\text{F}}=1}^{N_{\text{F}}}\frac{\|\hat{\boldsymbol{\mathcal{H}}}_{n_\text{F}}^{\text{CP}} - \boldsymbol{\mathcal{H}}_{n_\text{F}}^{\text{CP}}\|_{F}^{2}}{\|\boldsymbol{\mathcal{H}}_{n_\text{F}}^{\text{CP}}\|_{F}^{2}},
\end{equation}
where $\boldsymbol{\mathcal{H}}_{n_\text{F}}^{\text{CP}}$ and $\hat{\boldsymbol{\mathcal{H}}}_{n_\text{F}}^{\text{CP}}$ denote the groud-truth and predicted SFT domain channels of the upcoming OFDM symbols in the $n_\text{F}$-th frame, respectively. To ensure fairness, we evaluate all algorithms on the same QuaDRiGa-generated channel data.

\subsection{Computational Complexity Analysis}
The computational complexity of the proposed algorithm arises from tensor-matrix multiplications and scalar operations. The former is the main computational burden in the MO module (\alref{alg:MO_VI}) and perturbation parameter learning, while the latter dominates the SP module (\alref{alg:SP_VI}) and hyperparameter learning.
For tensor-matrix multiplications, the computational complexity of tensor-matrix multiplications in the horizontal antenna, vertical antenna, frequency, and temporal domains are given by $\mathcal{O}(C_{\text{h}})$, $\mathcal{O}(C_{\text{v}})$, $\mathcal{O}(C_{\text{sc}})$, and $\mathcal{O}(C_{\text{sym}})$, where $C_{\text{h}}$, $C_{\text{h}}$, $C_{\text{h}}$, and $C_{\text{h}}$ are defined by
\begin{subequations}
    \begin{equation}
        C_{\text{h}} {\;\triangleq\;} N_\text{h}K_\text{h}(K_\text{v}K_\text{de}K_\text{do} + N_\text{v}N_\text{sc}N_\text{sym}),
    \end{equation}
    \begin{equation}
        C_{\text{v}} {\;\triangleq\;} N_\text{v}K_\text{v}(N_\text{h}K_\text{de}K_\text{do} + K_\text{h}N_\text{sc}N_\text{sym}),
    \end{equation}
    \begin{equation}
        C_{\text{sc}} {\;\triangleq\;} N_\text{sc}K_\text{de}(N_\text{h}N_\text{v}K_\text{do} + K_\text{h}K_\text{v}N_\text{sym}),
    \end{equation}
    \begin{equation}
        C_{\text{sym}} {\;\triangleq\;} N_\text{sym}K_\text{do}(N_\text{h}N_\text{v}N_\text{sc} + K_\text{h}K_\text{v}K_\text{de}),
    \end{equation}
\end{subequations}
respectively.
When it comes to scalar operations, the computational complexity scales linearly with the size of ADD domain channels, which results in $\mathcal{O}(K_\text{h}K_\text{v}K_\text{de}K_\text{do})$ and is much smaller than that of tensor-matrix multiplications.
Therefore, the computational complexity of \alref{alg:Dynamic_CT} is expressed as $\mathcal{O}(T(C_{\text{h}} + C_{\text{v}} + C_{\text{sc}} + C_{\text{sym}}))$, where $T$ denotes the number of iterations.
To gain insight into the computational complexity reduction achieved by tensor-matrix multiplication compared to matrix-vector multiplication, we assume that the size of each dimension in the ADD domain channel is of the same order as that of the SFT domain channel, which typically holds in practical systems.
Therefore, the computational complexity can be further simplified to $\mathcal{O}(TN_\text{h}N_\text{v}N_\text{sc}N_\text{sym}(N_\text{h}+N_\text{v}+N_\text{sc} + N_\text{sym}))$, while for matrix-vector multiplication, ignoring the structure of separable factor matrices results in the computational complexity to become $\mathcal{O}(TN_\text{h}^{2}N_\text{v}^{2}N_\text{sc}^{2}N_\text{sym}^{2})$.

\begin{table}[!t]
  \caption{Computational Complexity}
  \label{tab:computational_complexity}
  \centering
  \begin{tabular}{cc}
  \toprule
  \textbf{Algorithm} & \textbf{Computational Complexity} \\
  \midrule
  Online TS-DCP & $\mathcal{O}(TN_\text{h}N_\text{v}N_\text{sc}N_\text{sym}(N_\text{h}+N_\text{v}+N_\text{sc} + N_\text{sym}))$ \\
  VKF &  $\mathcal{O}(P^{3}N_\text{h}^{3}N_\text{v}^{3}N_\text{sc})$ \\
  FIT &  $\mathcal{O}(TRN_\text{h}N_\text{v}N_\text{sc}N_\text{sym})$ \\
  PAD, MPAD &  $\mathcal{O}(N_\text{h}^{3}N_\text{v}^{3}N_\text{sc}N_\text{sym})$ \\
  \bottomrule
  \end{tabular}
\end{table}

In the bencmarks, the computational complexity of \textbf{PAD} and \textbf{MPAD} is dominated by the LMMSE algorithm for each pilot symbol and subcarrier, which is $\mathcal{O}(N_\text{h}^{3}N_\text{v}^{3}N_\text{sc}N_\text{sym})$. For \textbf{VKF},  the main contributor to computational complexity is the AR parameter learning, expressed as $\mathcal{O}(P^{3}N_\text{h}^{3}N_\text{v}^{3}N_\text{sc})$, where $P$ denotes the AR order, typically on the same order as the pilot OFDM symbols. 
In the case of \textbf{FIT}, its computational complexity is controlled by the ALS algorithm with $\mathcal{O}(TRN_\text{h}N_\text{v}N_\text{sc}N_\text{sym})$, where $T$ denotes the number of iterations and $R$ denotes the predefined tensor rank, typically set as the maximum number of possible paths in the environment. The computational complexity of the proposed algorithm and benchmarks are summarized in \tabref{tab:computational_complexity}.

\subsection{Performance Evaluation}

\subsubsection{TNMSE for Different Transmission Power}
\begin{figure}[!t]
  \centering
  \subfloat[]{
    \includegraphics[width = \linewidth]{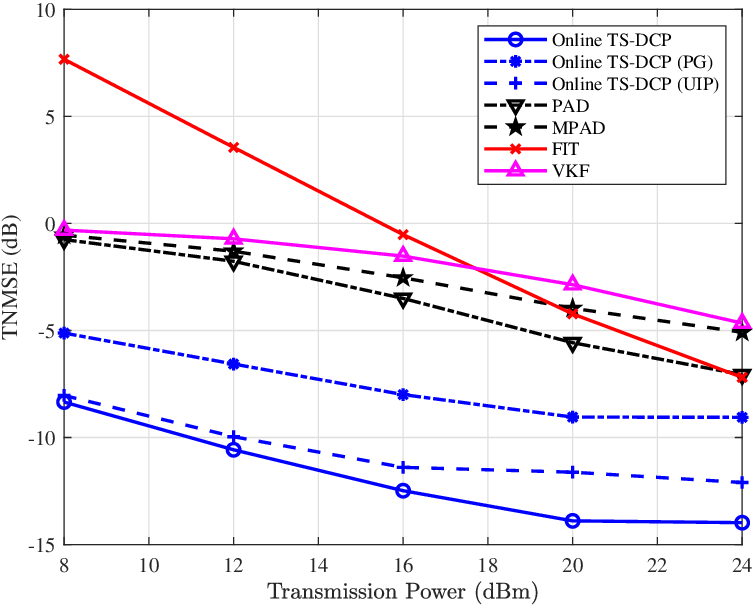}%
    \label{fig:PT_60kmph}
  }
  \\
  \subfloat[]{
    \includegraphics[width = \linewidth]{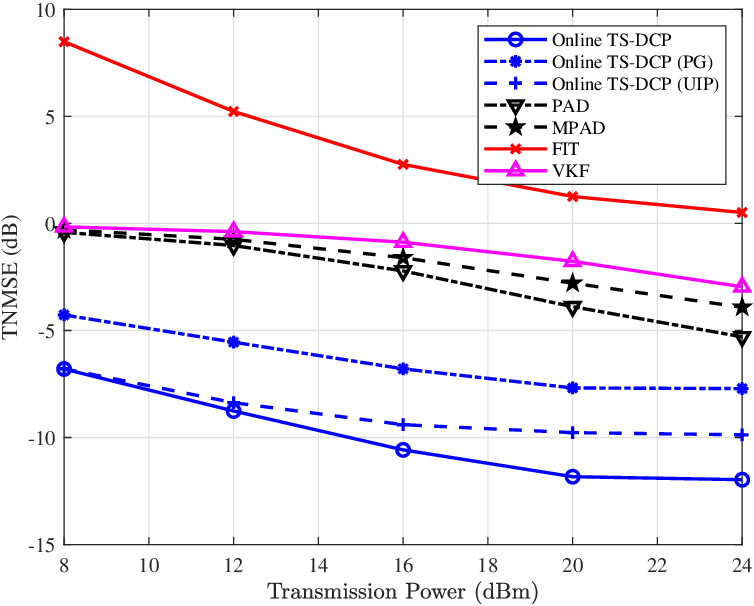}%
    \label{fig:PT_120kmph}
  }
  \caption{TNMSE of channel prediction versus transmission power: (a) $v=60$ km/h, (b) $v=120$ km/h.}
  \label{fig:PT}
\end{figure}

The TNMSE of the proposed algorithm and benchmarks versus transmission power for $v = 60$ km/h and $v = 120$ km/h are shown in \figref{fig:PT}. It is clear that the proposed algorithms significantly outperform the benchmarks across the full transmission power range.
Since \textbf{FIT} depends on the ALS algorithm, it struggles to incorporate statistical CSI and noise variance, resulting in a noticeable performance drop in the low transmission power regime.
Furthermore, \textbf{FIT} fails to achieve a TNMSE below $0$ dB at $v = 120$ km/h due to the inaccuracy of the first-order Taylor series in modeling temporal correlations at high mobility.
While \textbf{VKF} benefits from high-order AR processes and outperforms \textbf{FIT} in the low transmission power regime, it still struggles due to the neglect of inter-subcarrier and inter-antenna correlations. 
As a result, \textbf{VKF} fails to achieve a TNMSE below $-5$ dB at $P_\text{T} = 24$ dBm, even at $v = 60$ km/h.
By exploiting statistical CSI by LMMSE estimators and modeling temporal correlations with Prony and matrix pencil methods, \textbf{PAD} and \textbf{MPAD} outperform both \textbf{FIT} and \textbf{VKF} across the full transmission power range, with only marginal inferiority to \textbf{FIT} at $P_\text{T} {\;\geq\;} 20$ dBm and $v = 60$ km/h.
Among the proposed algorithms, \textbf{Online TS-DCP} is superior to both \textbf{Online TS-DCP (PG)} and \textbf{Online TS-DCP (UIP)}, owing to the perturbation parameter learning and structured prior.
Notably, the proposed algorithms still significantly outperform all benchmarks, even with pre-sampled grids and the unstructured independent prior, highlighting the improvement of channel prediction performance by Doppler domain modeling.

\begin{figure}[!t]
  \centering
  \subfloat[]{
    \includegraphics[width = \linewidth]{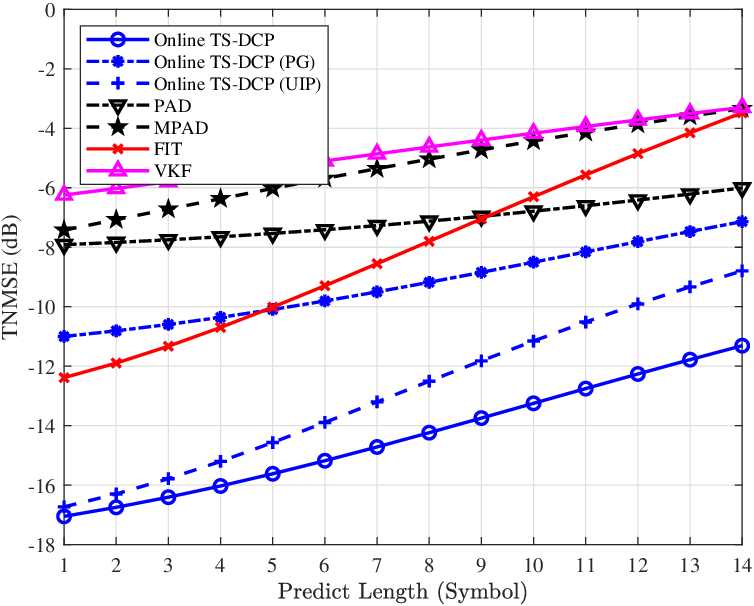}%
    \label{fig:PredictLength_60kmph_24dBm}
  }
  \\
  \subfloat[]{
    \includegraphics[width =  \linewidth]{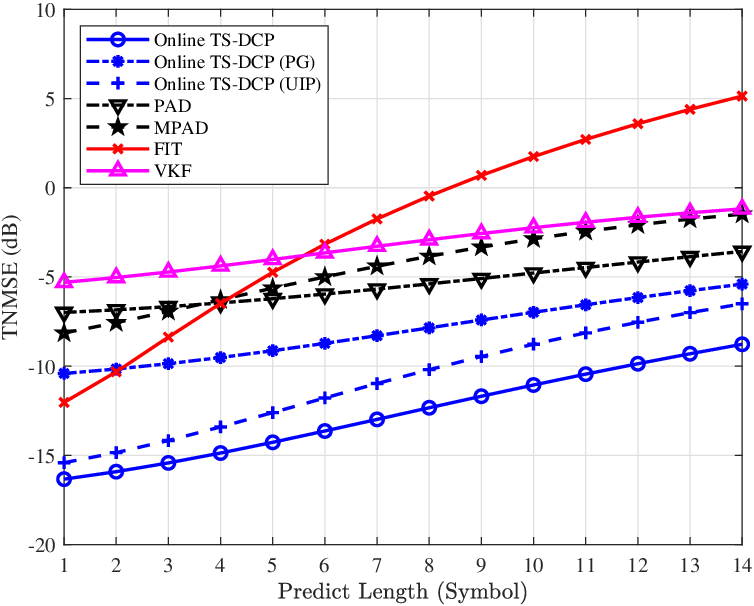}%
    \label{fig:PredictLength_120kmph_24dBm}
  }
  \caption{TNMSE of channel prediction versus predict length for $P_{\text{T}} = 24$ dBm: (a) $v=60$ km/h, (b) $v=120$ km/h.}
  \label{fig:PredictLength_24dBm}
\end{figure}

\subsubsection{TNMSE for Different Prediction Length}
To illustrate the variation of channel prediction performance with prediction length, we provide the performance for symbols between the current pilot symbols and the first future pilot symbols in \figref{fig:PredictLength_24dBm}.
As expected, the performance of the proposed algorithms and the benchmarks deteriorates with increasing prediction length, which is consistent with prior channel prediction studies.
Since \textbf{FIT} captures the temporal correlations by first-order Taylor series, the modeling error rises with the increasing prediction length, leading to an $8$ dB drop in TNMSE at $v = 60$ km/h and a $15$ dB drop at $v = 120$ km/h.
The TNMSE of \textbf{PAD} and \textbf{MPAD} is more stable with increasing prediction length, which suggests that Prony and matrix pencil methods provides a more accurate temporal correlation modeling than the first-order Taylor series.
The limited exploitation of inter-antenna and inter-subcarrier correlations hinders \textbf{VKF} from achieving satisfactory performance across all prediction lengths.
Due to the gridless algorithm in \textbf{FIT}, \textbf{Online TS-DCP (PG)} is slightly inferior to it for short prediction length, demonstrating the necessity of perturbation parameter learning.
Notably, \textbf{Online TS-DCP} and \textbf{Online TS-DCP (UIP)} outperform all benchmarks across the prediction lengths. 
Specifically, for the first future non-pilot symbols and pilot symbols, \textbf{Online TS-DCP} achieves TNMSEs below $-16$ dB and $-11$ dB at $v = 60$ km/h and below $-15$ dB and $-9$ dB at $v = 120$ km/h, which are unattainable by all other algorithms. 
The TNMSE gap between \textbf{Online TS-DCP (UIP)} and \textbf{Online TS-DCP} widens to more than $2$ dB with increasing prediction length, indicating that the structured prior contributes to the acquisition of more accurate ADD domain channels.

\subsubsection{CDF of Prediction Errors}
\begin{figure}[!t]
  \centering
  \subfloat[]{
    \includegraphics[width = \linewidth]{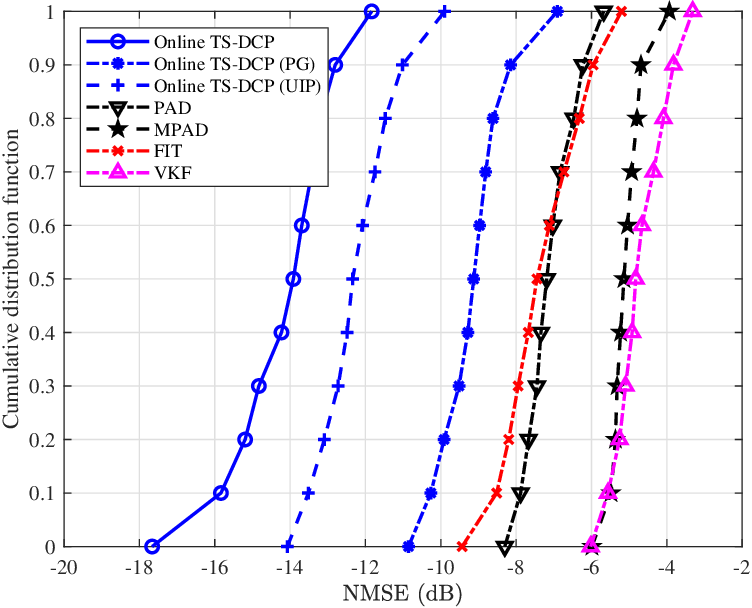}%
    \label{fig:CDF_60kmph_24dBm}
  }
  \\
  \subfloat[]{
    \includegraphics[width = \linewidth]{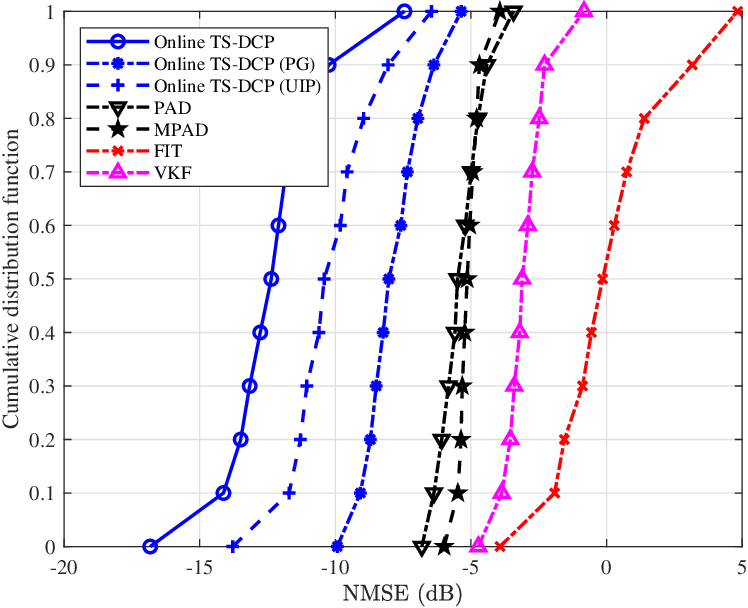}%
    \label{fig:CDF_120kmph_24dBm}
  }
  \caption{CDF of channel prediction error for $P_{\text{T}} = 24$ dBm: (a) $v=60$ km/h, (b) $v=120$ km/h.}
  \label{fig:CDF_24dBm}
\end{figure}

In \figref{fig:CDF_24dBm}, we present the cumulative distribution function (CDF) of the prediction errors, measured in terms of normalized mean square error (NMSE), for the proposed algorithms and benchmarks.
\textbf{FIT} and \textbf{PAD} exhibit similar prediction error distributions at $v  = 60$ km/h, while \textbf{FIT} is inferior to \textbf{PAD} at $v = 120$ km/h, but both outperform \textbf{VKF}.
This highlights the inaccuracy of the first-order Taylor series at high mobility and emphasizes the necessity of capturing correlations across antennas and subcarriers for accurate channel prediction.
Owing to the inherent similarity between Prony and matrix pencil methods, \textbf{PAD} and \textbf{MPAD} exhibit nearly identical error distributions, with a consistent performance gap observed under moderate mobility scenarios.
Furthermore, \textbf{Online TS-DCP (PG)}, \textbf{Online TS-DCP (UIP)}, and \textbf{Online TS-DCP} achieve the lowest prediction error for both mobility scenarios. 
Specifically, the $90$-th percentile of the prediction error for these algorithms is below $-12.5$ dB, $-11$ dB, and $-8$ dB at $v = 60$ km/h, and below $-10$ dB, $-7$ dB, and $-6$ dB at $v = 120$ km/h, respectively.

\subsubsection{Convergence Behaviors}
\begin{figure}[!t]
  \centering
  \subfloat[]{
    \includegraphics[width = \linewidth]{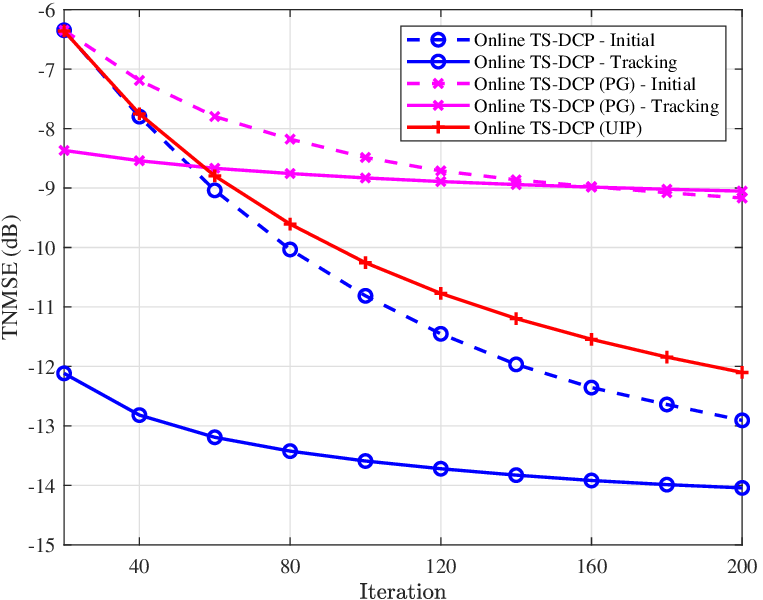}%
    \label{fig:Iteration_60kmph_24dBm}
  }
  \\
  \subfloat[]{
    \includegraphics[width = \linewidth]{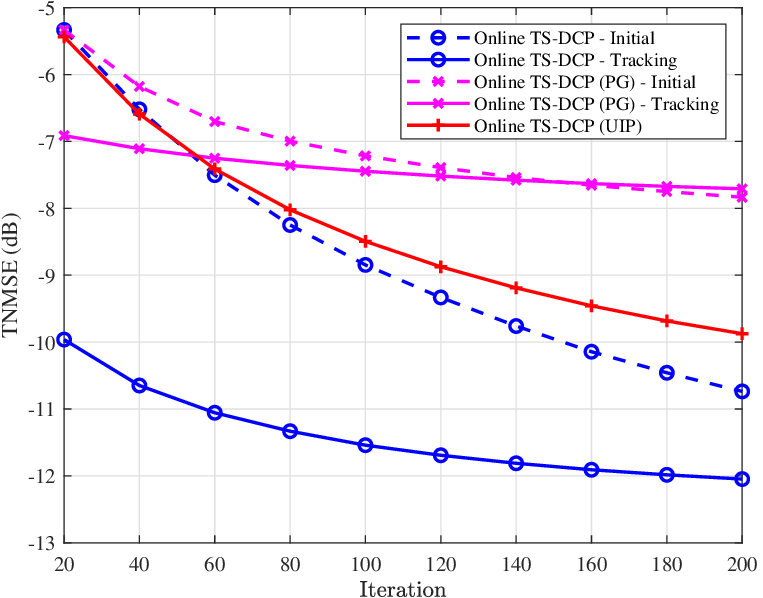}%
    \label{fig:Iteration_120kmph_24dBm}
  }
  \caption{The convergence performance of the proposed algorithms for $P_{\text{T}} = 24$ dBm: (a) $v=60$ km/h, (b) $v=120$ km/h.}
  \label{fig:Iteration_24dBm}
\end{figure}

To observe the trend of channel prediction performance with respect to the number of iterations, we present the convergence performance of \textbf{Online TS-DCP}, \textbf{Online TS-DCP (UIP)}, and \textbf{Online TS-DCP (PG)} at $P_\text{T} = 24$ dBm in \figref{fig:Iteration_24dBm}, separating the initial and tracking phases.
It is evident that \textbf{Online TS-DCP} and \textbf{Online TS-DCP (UIP)} exhibit fast convergence rates and achieve favorable convergence results in both phases, owing to the model accuracy improvement enabled by perturbation parameter learning.
Compared to \textbf{Online TS-DCP (PG)}, \textbf{Online TS-DCP} and \textbf{Online TS-DCP (UIP)} reduce the TNMSE from approximately $-9$ dB to nearly $-12$ dB and $-13$ dB at $v=60$ km/h, respectively, while from about $-8$ dB to below $-9.5$ dB and $-11$ dB at $v=120$ km/h, respectively.
Besides, the structured prior design in \textbf{Online TS-DCP} and \textbf{Online TS-DCP (PG)} enables a rapid reduction in TNMSE during the initial iterations of the tracking phase, allowing for early termination to reduce computational complexity based on practical system performance requirements.
Specifically, in the early iterations of the tracking phase, \textbf{Online TS-DCP} and \textbf{Online TS-DCP (PG)} achieve the TNMSE gains approximately $1.5{\sim}2$ dB and $4.5{\sim}5.5$ dB compared to the initial phase, respectively, with the gap narrowing as the number of iterations increases.

\section{Conclusion}\label{sec:conclusion}
This paper investigated TS-DCP for massive MIMO with temporal non-stationarity in moderate- to high-mobility scenarios.
Specifically, by framing the pilot symbols, we captured intra-frame and inter-frame correlations through Doppler domain modeling and Markov/AR processes, respectively.
Besides, we employed MRF and TCGD to model significant cross-domain correlations of the sparsity structure and power patterns in the ADD domain channel.
Building on the probabilistic models, the TS-DCP was formulated as the VFE minimization problem, unifying different variational inference rules through the structure design of the trial beliefs.
By leveraging the multi-linear structure of channels, the proposed online TS-DCP algorithm achieved substantial reductions in computational complexity.
Numerical results demonstrated significant superiority of proposed algorithms over benchmarks in channel prediction performance.
In the future, we will investigate deep learning approaches to further enhance the channel prediction performance in massive MIMO-OFDM systems under complicated propagation environments.

\appendices
\section{Proof of \propref{prop:online_approx}}\label{app:online_approx}
Based on the definition of KL divergence, the joint VFE in \eqref{eq:joint_VFE_minimization} is expressed by
\begin{equation}
    F_{\text{V}} = F_{\text{NE}} - F_{\text{EL}},
\end{equation}
where $F_{\text{NE}}$ and $F_{\text{EL}}$ denote the negative entropy and expected log-likelihood subfunctions, respectively, as given in \eqref{eq:joint_VFE_subfunction} at the top of the next page.
By substituting the constraints \eqref{eq:hyperpara_factor}, \eqref{eq:hyperpara_belief_HP}, and \eqref{eq:hyperpara_belief_ML} for the perturbation parameters and hyperparameters into the joint VFE, these subfunctions are simplified as shown in \eqref{eq:joint_VFE_subfunction_simplification} at the top of the next page.

To further decouple the joint VFE into per-frame VFEs, we factorize the joint PDF using the Bayes rule as \eqref{eq:joint_PDF_factorization} at the top of the next page.
By substituting the factorized PDF and trial belief in \eqref{eq:joint_PDF_factorization} and \eqref{eq:belief_online_approx} into the negative entropy and expected log-likelihood subfunctions, while incorporating normalization constraints of trial beliefs, we complete the proof.

\begin{figure*}[!t]
  \normalsize
  \begin{subequations}\label{eq:joint_VFE_subfunction}
    \begin{align}
        F_{\text{NE}} = \int \mathsf{b}( \boldsymbol{\mathcal{H}}_{(N_\text{F})}, \boldsymbol{\mathcal{G}}_{(N_\text{F})}, \boldsymbol{\mathcal{S}}_{(N_\text{F})}, \boldsymbol{\mathcal{Q}}_{(N_\text{F})}; \mathcal{P}_{\text{OG}, (N_\text{F})}, \mathcal{P}_{\text{HP}} ) \mathsf{ln} [\mathsf{b}( \boldsymbol{\mathcal{H}}_{(N_\text{F})}, \boldsymbol{\mathcal{G}}_{(N_\text{F})}, \boldsymbol{\mathcal{S}}_{(N_\text{F})}, \boldsymbol{\mathcal{Q}}_{(N_\text{F})}; \mathcal{P}_{\text{OG}, (N_\text{F})}, \mathcal{P}_{\text{HP}} )],
    \end{align}
    \begin{align}
        F_{\text{EL}} = \int \mathsf{b}( \boldsymbol{\mathcal{H}}_{(N_\text{F})}, \boldsymbol{\mathcal{G}}_{(N_\text{F})}, \boldsymbol{\mathcal{S}}_{(N_\text{F})}, \boldsymbol{\mathcal{Q}}_{(N_\text{F})}; \mathcal{P}_{\text{OG}, (N_\text{F})}, \mathcal{P}_{\text{HP}} ) \mathsf{ln} [\mathsf{P}( \boldsymbol{\mathcal{H}}_{(N_\text{F})}, \boldsymbol{\mathcal{G}}_{(N_\text{F})}, \boldsymbol{\mathcal{S}}_{(N_\text{F})}, \boldsymbol{\mathcal{Q}}_{(N_\text{F})}, \boldsymbol{\mathcal{Y}}_{(N_\text{F})}; \mathcal{P}_{\text{OG}, (N_\text{F})}, \mathcal{P}_{\text{HP}} )].
    \end{align}
  \end{subequations}
  \vspace*{-1em}
  \begin{subequations}\label{eq:joint_VFE_subfunction_simplification}
    \begin{align}
        F_{\text{NE}} = \int \mathsf{b}( \boldsymbol{\mathcal{H}}_{(N_\text{F})}, \boldsymbol{\mathcal{G}}_{(N_\text{F})}, \boldsymbol{\mathcal{S}}_{(N_\text{F})}, \boldsymbol{\mathcal{Q}}_{(N_\text{F})}) \mathsf{ln} [\mathsf{b}( \boldsymbol{\mathcal{H}}_{(N_\text{F})}, \boldsymbol{\mathcal{G}}_{(N_\text{F})}, \boldsymbol{\mathcal{S}}_{(N_\text{F})}, \boldsymbol{\mathcal{Q}}_{(N_\text{F})})],
    \end{align}
    \begin{align}
        F_{\text{EL}} = \int \mathsf{b}( \boldsymbol{\mathcal{H}}_{(N_\text{F})}, \boldsymbol{\mathcal{G}}_{(N_\text{F})}, \boldsymbol{\mathcal{S}}_{(N_\text{F})}, \boldsymbol{\mathcal{Q}}_{(N_\text{F})}) \mathsf{ln} [\mathsf{P}( \boldsymbol{\mathcal{H}}_{(N_\text{F})}, \boldsymbol{\mathcal{G}}_{(N_\text{F})}, \boldsymbol{\mathcal{S}}_{(N_\text{F})}, \boldsymbol{\mathcal{Q}}_{(N_\text{F})}, \boldsymbol{\mathcal{Y}}_{(N_\text{F})}; \hat{\mathcal{P}}_{\text{OG}, (N_\text{F})}, \hat{\mathcal{P}}_{\text{HP}} )].
    \end{align}
  \end{subequations}
  \hrulefill
\end{figure*}

\begin{figure*}[!t]
  \normalsize
  \begin{align}\label{eq:joint_PDF_factorization}
      \mathsf{P}( \boldsymbol{\mathcal{H}}_{(N_\text{F})}, &\boldsymbol{\mathcal{G}}_{(N_\text{F})}, \boldsymbol{\mathcal{S}}_{(N_\text{F})}, \boldsymbol{\mathcal{Q}}_{(N_\text{F})}, \boldsymbol{\mathcal{Y}}_{(N_\text{F})}; \hat{\mathcal{P}}_{\text{OG}, (N_\text{F})}, \hat{\mathcal{P}}_{\text{HP}} ) \nonumber \\
      &{\;\propto\;} \prod_{n_\text{F}=1}^{N_\text{F}} [\mathsf{P}( \boldsymbol{\mathcal{Y}}_{n_\text{F}} \mid \boldsymbol{\mathcal{H}}_{n_\text{F}}) \mathsf{P}( \boldsymbol{\mathcal{H}}_{n_\text{F}} \mid \boldsymbol{\mathcal{G}}_{n_\text{F}}; \hat{\mathcal{P}}_{\text{OG}, n_\text{F}}) \mathsf{P}(\boldsymbol{\mathcal{G}}_{n_\text{F}} \mid \boldsymbol{\mathcal{S}}_{n_\text{F}}, \boldsymbol{\mathcal{Q}}_{n_\text{F}}) \mathsf{P}( \boldsymbol{\mathcal{S}}_{n_\text{F}} \mid \boldsymbol{\mathcal{S}}_{n_\text{F} - 1}; \hat{\boldsymbol{\mathcal{M}}} ) \mathsf{P}( \boldsymbol{\mathcal{Q}}_{n_\text{F}} \mid \boldsymbol{\mathcal{Q}}_{n_\text{F} - 1}; \hat{\boldsymbol{\mathcal{L}}}, \hat{\boldsymbol{\mathcal{V}}} ) ] \nonumber \\
      &{\;\propto\;} \prod_{n_\text{F}=1}^{N_\text{F}} \mathsf{P} (\boldsymbol{\mathcal{H}}_{n_\text{F}}, \boldsymbol{\mathcal{G}}_{n_\text{F}}, \boldsymbol{\mathcal{S}}_{n_\text{F}}, \boldsymbol{\mathcal{Q}}_{n_\text{F}}, \boldsymbol{\mathcal{Y}}_{n_\text{F}} \mid \boldsymbol{\mathcal{S}}_{n_\text{F} - 1}, \boldsymbol{\mathcal{Q}}_{n_\text{F} - 1}; \hat{\mathcal{P}}_{\text{OG}, n_\text{F}}, \hat{\mathcal{P}}_{\text{HP}} )
  \end{align}
  \hrulefill
\end{figure*}

\section{Proof of \propref{prop:MO_FP}}\label{app:MO_FP}
The constrained BFE minimization problem can be solved by the Lagrange multiplier method \cite{boyd2004convex}. The Lagrange function is expressed as 
\begin{equation}
  L_{\text{MO}, n_\text{F}} = F_{\text{MO}, n_\text{F}} + L_{\text{MO-FO}, n_\text{F}} + L_{\text{MO-SO}, n_\text{F}},
\end{equation}
where $L_{\text{MO-FO}, n_\text{F}}$ and $L_{\text{MO-SO}, n_\text{F}}$ are the subfunctions of FO-SSCCs and SO-SSCCs defined in \eqref{eq:SubLagrange_MOFO} and \eqref{eq:SubLagrange_MOSO} at the top of the next page, respectively.

We begin with the fixed-point equations for factor beliefs $\mathsf{b}_{Y, n_\text{F}}$, $\mathsf{b}_{HG, n_\text{F}}$, and $\mathsf{b}_{G, n_\text{F}}$, given in \eqref{eq:fixed_point_b_Y}, \eqref{eq:fixed_point_b_HG}, and \eqref{eq:fixed_point_b_G}, respectively. By differentiating $L_{\text{MO}, n_\text{F}}$ with respect to $\mathsf{b}_{Y, n_\text{F}}$ and setting the result to zero, we can obtain the fixed-point equation of $\mathsf{b}_{Y, n_\text{F}}$, which satisfies
\begin{align}
    \mathsf{ln}(\mathsf{b}_{Y, n_\text{F}}) = & \mathsf{ln}(\mathsf{P}_{Y, n_\text{F}}) - 2\mathsf{Re}\{ \langle \boldsymbol{\mathcal{U}}_{n_\text{F}}^{H, b_{H}}, \boldsymbol{\mathcal{H}}_{n_\text{F}} \rangle \} + \nonumber \\
     \langle \boldsymbol{\mathcal{E}}_{n_\text{F}}^{H, b_{H}}, & \| \boldsymbol{\mathcal{H}}_{n_\text{F}} - \mathsf{E}_{\mathsf{b}_{Y, n_\text{F}}}\{ \boldsymbol{\mathcal{H}}_{n_\text{F}} \} \|^{{\odot}2} \rangle + C_{b_{Y}, n_\text{F}}, 
\end{align}
where $C_{b_{Y}, n_\text{F}}$ denotes the term independent of $\boldsymbol{\mathcal{H}}_{n_\text{F}}$. The fixed-point equation of $\mathsf{b}_{Y, n_\text{F}}$ is reformulated as in \eqref{eq:fixed_point_b_Y}, where the Gaussian distribution factor arises from the relaxation from MCCs to FO- and SO-SSCCs. 
Similarly, the fixed-point equations for $\mathsf{b}_{HG, n_\text{F}}$ and $\mathsf{b}_{G, n_\text{F}}$ given in \eqref{eq:fixed_point_b_HG} and \eqref{eq:fixed_point_b_G} can also be derived by differentiating $L_{\text{MO}, n_\text{F}}$ with respect to $\mathsf{b}_{HG, n_\text{F}}$ and $\mathsf{b}_{G, n_\text{F}}$, respectively, and setting the results to zero. In the fixed-point eqaution of $\mathsf{b}_{HG, n_\text{F}}$, we substitute the FO-SSCCs of $\mathsf{b}_{Y, n_\text{F}}$ and $\mathsf{b}_{HG, n_\text{F}}$ with respect to $\boldsymbol{\mathcal{H}}_{n_\text{F}}$, as well as $\mathsf{b}_{G, n_\text{F}}$ and $\mathsf{b}_{HG, n_\text{F}}$ with respect to $\boldsymbol{\mathcal{G}}_{n_\text{F}}$ to the derivatives.

Next, we move to the fixed-point equation for the beliefs of the variable beliefs $\mathsf{f}_{H, n_\text{F}}$ and $\mathsf{f}_{G, n_\text{F}}$, given by \eqref{eq:fixed_point_f_H} and \eqref{eq:fixed_point_f_G}, respectively. By differentiating $L_{\text{MO}, n_\text{F}}$ with respect to $\mathsf{f}_{H, n_\text{F}}$ and setting the result to zero, we obtain the fixed-point equation of $\mathsf{f}_{H, n_\text{F}}$, which satisfies
\begin{align}
    \ln(\mathsf{f}_{H, n_\text{F}}) = & \langle \boldsymbol{\mathcal{E}}_{n_\text{F}}^{H, b_{H}} +  \boldsymbol{\mathcal{E}}_{n_\text{F}}^{H, b_{HG}}, \| \boldsymbol{\mathcal{H}}_{n_\text{F}} - \mathsf{E}_{\mathsf{f}_{H, n_\text{F}}}\{ \boldsymbol{\mathcal{H}}_{n_\text{F}} \} \|^{{\odot}2} \rangle \nonumber \\
    - & 2\mathsf{Re}\{ \langle \boldsymbol{\mathcal{U}}_{n_\text{F}}^{H, b_{H}} + \boldsymbol{\mathcal{U}}_{n_\text{F}}^{H, b_{HG}}, \boldsymbol{\mathcal{H}}_{n_\text{F}} \rangle \} + C_{f_{H}, n_\text{F}}, 
\end{align}
where $C_{f_{H}, n_\text{F}}$ denotes the term independent of $\boldsymbol{\mathcal{H}}_{n_\text{F}}$. Due to the FO-SSCCs of $\mathsf{b}_{Y, n_\text{F}}$ and $\mathsf{f}_{H, n_\text{F}}$ with respect to $\boldsymbol{\mathcal{H}}_{n_\text{F}}$, the fixed-point equation of $\mathsf{f}_{H, n_\text{F}}$ is reformulated as in \eqref{eq:fixed_point_f_H}, where the product of Gaussian distribution factors arises from the relaxation from MCCs to FO- and SO-SSCCs of both $\mathsf{b}_{Y, n_\text{F}}$ and $\mathsf{b}_{HG, n_\text{F}}$. Similarly, the fixed-point equations for $\mathsf{f}_{G, n_\text{F}}$ given in \eqref{eq:fixed_point_f_G} can also be derived by differentiating $L_{\text{MO}, n_\text{F}}$ with respect to $\mathsf{f}_{G, n_\text{F}}$, substituting the FO-SSCCs of $\mathsf{b}_{G, n_\text{F}}$ and $\mathsf{f}_{G, n_\text{F}}$ with respect to $\boldsymbol{\mathcal{G}}_{n_\text{F}}$, and setting the result to zero.

\begin{figure*}[!t]
  \normalsize
  \begin{subequations}
    \begin{align}\label{eq:SubLagrange_MOFO}
      L_{\text{MO-FO}, n_\text{F}} =& 2 \mathsf{Re}\{ \langle \boldsymbol{\mathcal{U}}_{n_\text{F}}^{H, b_{H}}, [\mathsf{E}_{\mathsf{b}_{Y, n_\text{F}}}\{\boldsymbol{\mathcal{H}}_{n_\text{F}}\}  - \mathsf{E}_{\mathsf{f}_{H, n_\text{F}}}\{\boldsymbol{\mathcal{H}}_{n_\text{F}}\}] \rangle + \langle \boldsymbol{\mathcal{U}}_{n_\text{F}}^{H, b_{HG}}, [\mathsf{E}_{\mathsf{b}_{HG, n_\text{F}}}\{\boldsymbol{\mathcal{H}}_{n_\text{F}}\} - \mathsf{E}_{\mathsf{f}_{H, n_\text{F}}}\{\boldsymbol{\mathcal{H}}_{n_\text{F}}\}] \rangle + \nonumber \\
      & \langle \boldsymbol{\mathcal{U}}_{n_\text{F}}^{G, b_{G}}, [\mathsf{E}_{\mathsf{b}_{G, n_\text{F}}}\{\boldsymbol{\mathcal{G}}_{n_\text{F}}\} - \mathsf{E}_{\mathsf{f}_{G, n_\text{F}}}\{\boldsymbol{\mathcal{G}}_{n_\text{F}}\}] \rangle + \langle \boldsymbol{\mathcal{U}}_{n_\text{F}}^{G, b_{HG}}, [\mathsf{E}_{\mathsf{b}_{HG, n_\text{F}}}\{\boldsymbol{\mathcal{G}}_{n_\text{F}}\} - \mathsf{E}_{\mathsf{f}_{G, n_\text{F}}}\{\boldsymbol{\mathcal{G}}_{n_\text{F}}\}] \rangle \},
    \end{align}
    \vspace*{-1em}
    \begin{align}\label{eq:SubLagrange_MOSO}
      L_{\text{MO-SO}, n_\text{F}} = & \langle \boldsymbol{\mathcal{E}}_{n_\text{F}}^{H, b_{H}}, [\mathsf{V}_{\mathsf{b}_{Y, n_\text{F}}}\{\boldsymbol{\mathcal{H}}_{n_\text{F}}\}  - \mathsf{V}_{\mathsf{f}_{H, n_\text{F}}}\{\boldsymbol{\mathcal{H}}_{n_\text{F}}\}] \rangle + \langle \boldsymbol{\mathcal{E}}_{n_\text{F}}^{H, b_{HG}}, [\mathsf{V}_{\mathsf{b}_{HG, n_\text{F}}}\{\boldsymbol{\mathcal{H}}_{n_\text{F}}\} - \mathsf{V}_{\mathsf{f}_{H, n_\text{F}}}\{\boldsymbol{\mathcal{H}}_{n_\text{F}}\}] \rangle  \nonumber \\
      & \langle \boldsymbol{\mathcal{E}}_{n_\text{F}}^{G, b_{G}}, [\mathsf{V}_{\mathsf{b}_{G, n_\text{F}}}\{\boldsymbol{\mathcal{G}}_{n_\text{F}}\} - \mathsf{V}_{\mathsf{f}_{G, n_\text{F}}}\{\boldsymbol{\mathcal{G}}_{n_\text{F}}\}] \rangle + \langle \boldsymbol{\mathcal{E}}_{n_\text{F}}^{G, b_{HG}}, [\mathsf{V}_{\mathsf{b}_{HG, n_\text{F}}}\{\boldsymbol{\mathcal{G}}_{n_\text{F}}\} - \mathsf{V}_{\mathsf{f}_{G, n_\text{F}}}\{\boldsymbol{\mathcal{G}}_{n_\text{F}}\}] \rangle. 
    \end{align}
  \end{subequations}
  \hrulefill
  \vspace*{4pt}
\end{figure*}

\section{Proof of \propref{prop:perturbation}}\label{app:perturbation}
Due to the singularity in Dirac Delta functions, the conditional PDF $\mathsf{P}(\boldsymbol{\mathcal{H}}_{n_\text{F}}\!\mid\!\boldsymbol{\mathcal{G}}_{n_\text{F}}; \hat{\mathcal{P}}_{{\backslash}\text{h}, n_\text{F}})$ is approximated by the Gaussian distribution with variance approaching zero. On this basis, the objective function for the perturbation parameter learning rule in the horizontal angle domain is expressed as
\begin{equation}
  f(\hat{\bm{\Delta}}{\theta}_{n_\text{F}}) = f_{1}(\hat{\bm{\Delta}}{\theta}_{n_\text{F}}) + f_{2}(\hat{\bm{\Delta}}{\theta}_{n_\text{F}}),
\end{equation}
where $f_{1}(\hat{\bm{\Delta}}{\theta}_{n_\text{F}})$ and $f_{2}(\hat{\bm{\Delta}}{\theta}_{n_\text{F}})$ denote the objective subfunctions, defined by \eqref{eq:Objective_Subfunction} at the top of the next page. 

The parametrized factor matrix ${\mathbf{A}}_{\text{h}}(\bar{\bm{\theta}} + \hat{\bm{\Delta}}\bm{\theta}_{n_\text{F}})$ is approximated by the first-order Taylor series, given by
\begin{equation}
  {\mathbf{A}}_{\text{h}}(\bar{\bm{\theta}} + \hat{\bm{\Delta}}\bm{\theta}_{n_\text{F}}) {\;\approx\;} {\mathbf{A}}_{\text{h}}(\bar{\bm{\theta}}) + \dot{\mathbf{A}}_{\text{h}}(\bar{\bm{\theta}})\mathsf{diag}\{ \hat{\bm{\Delta}}\bm{\theta}_{n_\text{F}} \},
\end{equation}
where $\dot{\mathbf{A}}_{\text{h}}(\cdot)$ denotes the derivative matrix of the horizontal angle domain parametrized factor matrix with respect to the horizontal direction cosine. 
Based on the Taylor approximation and the commutativity of tensor-matrix multiplication across dimensions, the first subfunction $f_{1}(\hat{\bm{\Delta}}{\theta}_{n_\text{F}})$ is expressed as
\begin{align}
  f_{1}(\hat{\bm{\Delta}}{\theta}_{n_\text{F}}) = & \| {\Delta}\hat{\boldsymbol{\mathcal{H}}}_{n_\text{F}} - \hat{\boldsymbol{\mathcal{G}}}_{n_\text{F}}^{\text{h}} {\;\times_{1}\;} \dot{\mathbf{A}}_{\text{h}}(\bar{\bm{\theta}})\mathsf{diag}\{ \hat{\bm{\Delta}}\bm{\theta}_{n_\text{F}} \} \|_{F}^{2} \nonumber \\
  = & \sum_{n} \| {\Delta}\hat{\mathbf{h}}_{n_\text{F}, n}^{(1)} - \dot{\mathbf{A}}_{\text{h}}(\bar{\bm{\theta}}) \mathsf{diag}\{ \mathbf{g}_{n_\text{F}, n}^{\text{h}} \} \hat{\bm{\Delta}}\bm{\theta}_{n_\text{F}} \|_{F}^{2} \nonumber \\
  = & \hat{\bm{\Delta}}\bm{\theta}_{n_\text{F}}^{T} \bm{\Pi}_{\text{h}, 1} \hat{\bm{\Delta}}\bm{\theta}_{n_\text{F}} - 2\bm{\mu}_{\text{h}, 1}^{T}\hat{\bm{\Delta}}\bm{\theta}_{n_\text{F}} + C_{1},
\end{align}
where ${\Delta}\hat{\boldsymbol{\mathcal{H}}}_{n_\text{F}}$, 
$\hat{\boldsymbol{\mathcal{G}}}_{n_\text{F}}^{\text{h}}$, ${\Delta}\hat{\mathbf{h}}_{n_\text{F}, n}$, and $\mathbf{g}_{n_\text{F}, n}^{\text{h}}$ are defined in this proposition, $C_{1}$ denotes the term independent of $\hat{\bm{\Delta}}{\theta}_{n_\text{F}}$, $\bm{\Pi}_{\text{h}, 1}$ and $\bm{\mu}_{\text{h}, 1}$ denote the auxiliary matrix and vector, given by
\begin{subequations}
  \begin{equation}
    \bm{\Pi}_{\text{h}, 1} = \sum_{n}  (\dot{\mathbf{A}}_{\text{h}}^{H}(\bar{\bm{\theta}})\dot{\mathbf{A}}_{\text{h}}(\bar{\bm{\theta}}))^{\ast} {\;\odot\;} (\hat{\mathbf{g}}_{n_\text{F}, n}\hat{\mathbf{g}}_{n_\text{F}, n}^{H}),
  \end{equation}
  \begin{equation}
    \bm{\mu}_{\text{h}, 1} = \sum_{n} \mathsf{Re}\{ \mathsf{diag}^{H}\{ \mathbf{g}_{n_\text{F}, n}^{\text{h}} \} \dot{\mathbf{A}}_{\text{h}}^{H}(\bar{\bm{\theta}}) {\Delta}\hat{\mathbf{h}}_{n_\text{F}, n}^{(1)} \}.
  \end{equation}
\end{subequations}

Similarly, the second subfunction $f_{2}(\hat{\bm{\Delta}}{\theta}_{n_\text{F}})$ is expressed as
\begin{align}
  f_{2}(\hat{\bm{\Delta}}{\theta}_{n_\text{F}}) =& \| \boldsymbol{\mathcal{E}}_{G, n_\text{F}}^{\text{h}} {\;\times_{1}\;} | \tilde{\mathbf{A}}_{\text{h}}(\bar{\bm{\theta}} + \hat{\bm{\Delta}}\bm{\theta}_{n_\text{F}}) |^{{\odot}2} \|_{1} \nonumber \\
  =& \sum_{n} \| \tilde{\mathbf{A}}_{\text{h}}(\bar{\bm{\theta}} + \hat{\bm{\Delta}}\bm{\theta}_{n_\text{F}}) |^{{\odot}2} \bm{\varepsilon}_{G, n_\text{F}, n}^{\text{h}} \|_{1} \nonumber \\
  =& \hat{\bm{\Delta}}\bm{\theta}_{n_\text{F}}^{T} \bm{\Pi}_{\text{h}, 2} \hat{\bm{\Delta}}\bm{\theta}_{n_\text{F}} - 2\bm{\mu}_{\text{h}, 2}^{T}\hat{\bm{\Delta}}\bm{\theta}_{n_\text{F}} + C_{2},
\end{align}
where $\boldsymbol{\mathcal{E}}_{G, n_\text{F}}^{\text{h}}$ and $\bm{\varepsilon}_{G, n_\text{F}, n}^{\text{h}}$ are defined in this proposition, $C_{2}$ denotes the term independent of $\hat{\bm{\Delta}}{\theta}_{n_\text{F}}$, $\bm{\Pi}_{\text{h}, 2}$ and $\bm{\mu}_{\text{h}, 2}$ denote the auxiliary matrix and vector, given by
\begin{subequations}
  \begin{equation}
    \bm{\Pi}_{\text{h}, 2} = \sum_{n} (\dot{\mathbf{A}}_{\text{h}}^{H}(\bar{\bm{\theta}})\dot{\mathbf{A}}_{\text{h}}(\bar{\bm{\theta}}))^{\ast} {\;\odot\;} \mathsf{diag}\{ \bm{\varepsilon}_{G, n_\text{F}, n}^{\text{h}} \} ,
  \end{equation}
  \begin{equation}
    \bm{\mu}_{\text{h}, 2} = - \sum_{n} \mathsf{Re}\{ \mathsf{diag}\{ \dot{\mathbf{A}}_{\text{h}}^{H}(\bar{\bm{\theta}}) \tilde{\mathbf{A}}_{\text{h}}(\bar{\bm{\theta}})\} {\;\odot\;} \bm{\varepsilon}_{G, n_\text{F}, n}^{\text{h}} \}.
  \end{equation}
\end{subequations}
This concludes the proof.

\begin{figure*}[!t]
  \normalsize
  \begin{subequations}\label{eq:Objective_Subfunction}
    \begin{equation}
      f_{1} (\hat{\bm{\Delta}}{\theta}_{n_\text{F}}) = \| \hat{\boldsymbol{\mathcal{H}}}_{n_\text{F}} - \hat{\boldsymbol{\mathcal{G}}}_{n_\text{F}} {\;\times_{1}\;} \tilde{\mathbf{A}}_{\text{h}}(\bar{\bm{\theta}} + \hat{\bm{\Delta}}\bm{\theta}_{n_\text{F}}) {\;\times_{2}\;} {\mathbf{A}}_{\text{v}}(\bar{\bm{\phi}} + \hat{\bm{\Delta}}{\bm{\phi}}_{n_\text{F}}^{\star}) {\;\times_{3}\;} {\mathbf{B}}(\bar{\bm{\tau}} + \hat{\bm{\Delta}}{\bm{\tau}}_{n_\text{F}}^{\star}) {\;\times_{4}\;} {\mathbf{C}}(\bar{\bm{\nu}} + \hat{\bm{\Delta}}{\bm{\nu}}_{n_\text{F}}^{\star}) \|_{F}^{2},
    \end{equation}
    \begin{equation}
      f_{2} (\hat{\bm{\Delta}}{\theta}_{n_\text{F}}) = \| \boldsymbol{\mathcal{E}}_{G, n_\text{F}} {\;\times_{1}\;} |\tilde{\mathbf{A}}_{\text{h}}(\bar{\bm{\theta}} + \hat{\bm{\Delta}}\bm{\theta}_{n_\text{F}})|^{{\odot}2} {\;\times_{2}\;} |{\mathbf{A}}_{\text{v}}(\bar{\bm{\phi}} + \hat{\bm{\Delta}}{\bm{\phi}}_{n_\text{F}}^{\star})|^{{\odot}2} {\;\times_{3}\;} |{\mathbf{B}}(\bar{\bm{\tau}} + \hat{\bm{\Delta}}{\bm{\tau}}_{n_\text{F}}^{\star})|^{{\odot}2} {\;\times_{4}\;} |{\mathbf{C}}(\bar{\bm{\nu}} + \hat{\bm{\Delta}}{\bm{\nu}}_{n_\text{F}}^{\star})|^{{\odot}2} \|_{1},
    \end{equation}
  \end{subequations}
  \hrulefill
  \vspace*{4pt}
\end{figure*}

\section{Proof of \propref{prop:hyperparameter}}\label{app:hyperparameter}
We begin with the learning rule for the transition factor $\hat{\boldsymbol{\mathcal{M}}}$ in the probabilistic model of hidden state tensors. The objective function of VFE minimization problem with respect to $\hat{\boldsymbol{\mathcal{M}}}$ in \eqref{eq:learning_rule_M} can be decomposed into the summation of $N_\text{F}$ subfunctions, given by
\begin{equation}
    f_{\text{HS}, \text{TC}}(\hat{\boldsymbol{\mathcal{M}}}) = \sum_{n_\text{F} = 1}^{N_\text{F}} f_{\text{HS}, \text{TC}, n_\text{F}}(\hat{\boldsymbol{\mathcal{M}}}),
\end{equation}
where $f_{\text{HS}, \text{TC}}(\hat{\boldsymbol{\mathcal{M}}})$ denotes the objective function in \eqref{eq:learning_rule_M} and $f_{\text{HS}, \text{TC}, n_\text{F}}(\hat{\boldsymbol{\mathcal{M}}})$ denotes the $n_\text{F}$-th subfunction, given by
\begin{equation}
    f_{\text{HS}, \text{TC}, n_\text{F}}(\hat{\boldsymbol{\mathcal{M}}}) = \mathsf{E}_{\mathsf{b}_{S, n_\text{F}}} \{ \ln \mathsf{P}_{\text{TC}} (\boldsymbol{\mathcal{S}}_{n_\text{F}} \mid \boldsymbol{\mathcal{S}}_{n_\text{F} - 1}; \hat{\boldsymbol{\mathcal{M}}}) \}.
\end{equation}
Since the normalization coefficients in the temporal correlation probabilistic model involve the hyperparameter $\hat{\boldsymbol{\mathcal{M}}}$, the exact expression of the probabilistic model is given by
\begin{align}
    \mathsf{P}_{\text{TC}} & (\boldsymbol{\mathcal{S}}_{n_\text{F}} \mid \boldsymbol{\mathcal{S}}_{n_\text{F} - 1}; \hat{\boldsymbol{\mathcal{M}}}) = \nonumber \\
    &C_{\text{TC}, n_\text{F}}^{-1} \mathsf{exp}(\langle \hat{\boldsymbol{\mathcal{M}}}, (2 \boldsymbol{\mathcal{S}}_{n_\text{F} - 1} - 1) {\;\odot\;} (2 \boldsymbol{\mathcal{S}}_{n_\text{F}} - 1) \rangle),
\end{align}
where $C_{\text{TC}, n_\text{F}}$ denotes the normalization coefficient, given by
\begin{align}
    C_{\text{TC}, n_\text{F}} = & 2^{K_\text{h}K_\text{v}K_\text{de}K_\text{do}} \prod_{k_\text{h}, k_\text{v}, k_\text{de}, k_\text{do}} \mathsf{exp}(-[\hat{\boldsymbol{\mathcal{M}}}]_{k_\text{h}, k_\text{v}, k_\text{de}, k_\text{do}}) \nonumber \\
    & \prod_{k_\text{h}, k_\text{v}, k_\text{de}, k_\text{do}} [ \mathsf{exp}(2[\hat{\boldsymbol{\mathcal{M}}}]_{k_\text{h}, k_\text{v}, k_\text{de}, k_\text{do}}) + 1].
\end{align}
Therefore, the derivative of the $n_\text{F}$-th subfunction with respect to $[\hat{\boldsymbol{\mathcal{M}}}]_{k_\text{h}, k_\text{v}, k_\text{de}, k_\text{do}}$ is given by
\begin{align}
    &\frac{{\partial} f_{\text{HS}, \text{TC}, n_\text{F}}(\hat{\boldsymbol{\mathcal{M}}}) }{{\partial} [\hat{\boldsymbol{\mathcal{M}}}]_{k_\text{h}, k_\text{v}, k_\text{de}, k_\text{do}} } {\;\approx\;} \frac{1 - \mathsf{exp}(2[\hat{\boldsymbol{\mathcal{M}}}]_{k_\text{h}, k_\text{v}, k_\text{de}, k_\text{do}})}{1 + \mathsf{exp}(2[\hat{\boldsymbol{\mathcal{M}}}]_{k_\text{h}, k_\text{v}, k_\text{de}, k_\text{do}})} + \nonumber \\
    &{\qquad} (2 [\hat{\boldsymbol{\mathcal{S}}}_{n_\text{F}}]_{k_\text{h}, k_\text{v}, k_\text{de}, k_\text{do}} - 1)(2 [\hat{\boldsymbol{\mathcal{S}}}_{n_\text{F} - 1}]_{k_\text{h}, k_\text{v}, k_\text{de}, k_\text{do}} - 1),
\end{align}
where the approximation follows from the online approximation, in which the hidden state tensors from the previous frame are approximated using the posterior mean.
By summing the derivative of all objective subfunctions and setting the result to zero, the learning rule for $[\hat{\boldsymbol{\mathcal{M}}}]_{k_\text{h}, k_\text{v}, k_\text{de}, k_\text{do}}$ is expressed as
\begin{equation}
    [\hat{\boldsymbol{\mathcal{M}}}^{\star}]_{k_\text{h}, k_\text{v}, k_\text{de}, k_\text{do}} = \ln \frac{1 + [\hat{\boldsymbol{\mathcal{K}}}_{M}]_{k_\text{h}, k_\text{v}, k_\text{de}, k_\text{do}}}{1 - [\hat{\boldsymbol{\mathcal{K}}}_{M}]_{k_\text{h}, k_\text{v}, k_\text{de}, k_\text{do}}}, 
\end{equation}
where $[\hat{\boldsymbol{\mathcal{K}}}_{M}]_{k_\text{h}, k_\text{v}, k_\text{de}, k_\text{do}}$ is defined by
\begin{align}
    [\hat{\boldsymbol{\mathcal{K}}}_{M}&]_{k_\text{h}, k_\text{v}, k_\text{de}, k_\text{do}} = \\
    \frac{1}{N_\text{F}} &\sum_{n_\text{F} = 1}^{N_\text{F}} (2 [\hat{\boldsymbol{\mathcal{S}}}_{n_\text{F}}]_{k_\text{h}, k_\text{v}, k_\text{de}, k_\text{do}} - 1)(2 [\hat{\boldsymbol{\mathcal{S}}}_{n_\text{F} - 1}]_{k_\text{h}, k_\text{v}, k_\text{de}, k_\text{do}} - 1). \nonumber
\end{align}
The derivation of the learning rule for transition factors can be completed by organizing the above results into tensor form.

Similar to the case of transition factors $\hat{\boldsymbol{\mathcal{M}}}$, the objective function of VFE minimization problem with respect to the temporal correlation factor $\hat{\boldsymbol{\mathcal{L}}}$ can be decomposed into the summation of $N_\text{F}$ subfunctions, given by
\begin{equation}
    f_{\text{HV}, \text{TC}}(\hat{\boldsymbol{\mathcal{L}}}) = \sum_{n_\text{F} = 1}^{N_\text{F}} f_{\text{HV}, \text{TC}, n_\text{F}}(\hat{\boldsymbol{\mathcal{L}}}),
\end{equation}
where $f_{\text{HV}, \text{TC}}(\hat{\boldsymbol{\mathcal{L}}})$ denotes the objective function in \eqref{eq:learning_rule_LV} and $f_{\text{HV}, \text{TC}, n_\text{F}}(\hat{\boldsymbol{\mathcal{L}}})$ denotes the $n_\text{F}$-th subfunction, given by
\begin{equation}
    f_{\text{HV}, \text{TC}, n_\text{F}}(\hat{\boldsymbol{\mathcal{L}}}) = \mathsf{E}_{\mathsf{b}_{Q, n_\text{F}}} \{ \ln \mathsf{P} (\boldsymbol{\mathcal{Q}}_{n_\text{F}} \mid \boldsymbol{\mathcal{Q}}_{n_\text{F} - 1}; \hat{\boldsymbol{\mathcal{L}}}, \hat{\boldsymbol{\mathcal{V}}}) \},
\end{equation}
where the temporal correlation model $\mathsf{P} (\boldsymbol{\mathcal{Q}}_{n_\text{F}} \mid \boldsymbol{\mathcal{Q}}_{n_\text{F} - 1}; \hat{\boldsymbol{\mathcal{L}}}, \hat{\boldsymbol{\mathcal{V}}})$ is expressed as
\begin{align}
    & \mathsf{P} ( \boldsymbol{\mathcal{Q}}_{n_\text{F}} \mid  \boldsymbol{\mathcal{Q}}_{n_\text{F} - 1}; \hat{\boldsymbol{\mathcal{L}}}, \hat{\boldsymbol{\mathcal{V}}}) {\;\propto\;} \nonumber \\
    & \begin{cases} \mathsf{CN}(\boldsymbol{\mathcal{Q}}_{n_\text{F}}; (1 - \hat{\boldsymbol{\mathcal{L}}}){\;\odot\;}\boldsymbol{\mathcal{Q}}_{n_\text{F} - 1}, \hat{\boldsymbol{\mathcal{L}}}^{{\odot}2} {\;\odot\;} \bar{\boldsymbol{\mathcal{V}}}), & n_\text{F} > 1 \\
    \mathsf{CN}(\boldsymbol{\mathcal{Q}}_{n_\text{F}}; \boldsymbol{\mathcal{C}}(0), \bar{\boldsymbol{\mathcal{V}}}), & n_\text{F} = 1
    \end{cases},
\end{align}
where $\bar{\boldsymbol{\mathcal{V}}}$ is defined in \eqref{eq:equivalent_V}. It can be observed that $f_{\text{HV}, \text{TC}, 1}(\hat{\boldsymbol{\mathcal{L}}})$ is independent of $\hat{\boldsymbol{\mathcal{L}}}$, meaning that only the subfunctions with $n_\text{F} > 1$ need to be considered. By differentiating the subfunctions with $n_\text{F} > 1$ with respect to $\hat{\boldsymbol{\mathcal{L}}}$, we yield
\begin{align}
    &-\frac{{\partial} f_{\text{HV}, \text{TC}, n_\text{F}}(\hat{\boldsymbol{\mathcal{L}}})}{{\partial} \hat{\boldsymbol{\mathcal{L}}}} {\;\approx\;} \nonumber \\
    &{\quad}2[\bar{\boldsymbol{\mathcal{V}}} {\;\odot\;} \hat{\boldsymbol{\mathcal{L}}}^{{\odot}2} + \boldsymbol{\mathcal{K}}_{L, 1, n_\text{F}} {\;\odot\;} \hat{\boldsymbol{\mathcal{L}}} + \boldsymbol{\mathcal{K}}_{L, 0, n_\text{F}}] {\;\oslash\;} (\hat{\boldsymbol{\mathcal{L}}}^{{\odot}3}{\;\odot\;}\bar{\boldsymbol{\mathcal{V}}}),
\end{align}
where the approximation also follows from the online approximation, $ \boldsymbol{\mathcal{K}}_{L, 1, n_\text{F}}$ and $ \boldsymbol{\mathcal{K}}_{L, 0, n_\text{F}}$ are defined as
\begin{subequations}
    \begin{equation}
          \boldsymbol{\mathcal{K}}_{L, 1, n_\text{F}}{\;\triangleq\;} | \hat{\boldsymbol{\mathcal{Q}}}_{n_\text{F}-1} |^{{\odot}2} - \mathsf{Re}\{ \hat{\boldsymbol{\mathcal{Q}}}_{n_\text{F}-1}^{\ast} {\;\odot\;} \hat{\boldsymbol{\mathcal{Q}}}_{n_\text{F}} \} , 
        \end{equation}
        \begin{equation}
          \boldsymbol{\mathcal{K}}_{L, 0, n_\text{F}} {\;\triangleq\;} - |\hat{\boldsymbol{\mathcal{Q}}}_{n_\text{F}} - \hat{\boldsymbol{\mathcal{Q}}}_{n_\text{F}-1}|^{{\odot}2} - \boldsymbol{\mathcal{E}}_{Q, n_\text{F}}.
        \end{equation}
\end{subequations}
The derivation of the learning rule for temporal correlation factor is completed by summing the derivatives of the subfunctions with $n_\text{F} > 1$ and enforcing the non-zero constraints on both $\hat{\boldsymbol{\mathcal{L}}}$ and $\bar{\boldsymbol{\mathcal{V}}}$, which concludes the proof.



\ifCLASSOPTIONcaptionsoff
  \newpage
\fi



\bibliographystyle{IEEEtran}
\bibliography{IEEEabrv, reference}
%

%








\end{document}